%% file: main.tex
\pdfoutput=1
\documentclass[12pt,a4paper]{article}

\usepackage{ifthen} 
\newboolean{pdflatex}
\setboolean{pdflatex}{true} 

\newboolean{articletitles}
\setboolean{articletitles}{true} 

\newboolean{uprightparticles}
\setboolean{uprightparticles}{false} 

\def\paperauthors{LHCb collaboration} 
\def\paperasciititle{Search for CP violation in D(s)+ -> h+ pi0 and D(s)+ -> h+ eta decays} 
\def\papertitle{Search for \CP violation in $\decay{D_{(s)}^{+}}{h^{+}\piz}$ and $\decay{D_{(s)}^{+}}{h^{+}\eta}$ decays} 
\def\paperkeywords{{High Energy Physics}, {LHCb}} 
\def\papercopyright{\the\year\ CERN for the benefit of the LHCb collaboration} 
\def\paperlicence{CC BY 4.0 licence}
\def\paperlicenceurl{https://creativecommons.org/licenses/by/4.0/}

\input{preamble}
\usepackage{longtable} 
\usepackage{subcaption} 
\usepackage{multirow} 
\usepackage{booktabs} 
\begin{document}

\renewcommand{\thefootnote}{\fnsymbol{footnote}}
\setcounter{footnote}{1}

\input{title-LHCb-PAPER}


\renewcommand{\thefootnote}{\arabic{footnote}}
\setcounter{footnote}{0}

\cleardoublepage


\pagestyle{plain} 
\setcounter{page}{1}
\pagenumbering{arabic}


\input{body}

\input{acknowledgements}




\addcontentsline{toc}{section}{References}
\bibliographystyle{LHCb}
\bibliography{main,standard,LHCb-PAPER,LHCb-CONF,LHCb-DP,LHCb-TDR}

\newpage
\input{Authorship_LHCb-PAPER-2021-001}

\end{document}

%% file: preamble.tex
\usepackage[top=1in, bottom=1.25in, left=1in, right=1in]{geometry}

\usepackage{microtype}
\usepackage{lineno}  
\usepackage{xspace} 
\usepackage{caption} 

\usepackage{graphicx}  
\usepackage{color}
\usepackage{colortbl}
\graphicspath{{./figs/}} 

\usepackage{amsmath} 
\usepackage{amssymb}
\usepackage{amsfonts}
\usepackage{upgreek} 

\newcommand*\patchAmsMathEnvironmentForLineno[1]{%
\expandafter\let\csname old#1\expandafter\endcsname\csname #1\endcsname
\expandafter\let\csname oldend#1\expandafter\endcsname\csname
end#1\endcsname
 \renewenvironment{#1}%
   {\linenomath\csname old#1\endcsname}%
   {\csname oldend#1\endcsname\endlinenomath}%
}
\newcommand*\patchBothAmsMathEnvironmentsForLineno[1]{%
  \patchAmsMathEnvironmentForLineno{#1}%
  \patchAmsMathEnvironmentForLineno{#1*}%
}
\AtBeginDocument{%
\patchBothAmsMathEnvironmentsForLineno{equation}%
\patchBothAmsMathEnvironmentsForLineno{align}%
\patchBothAmsMathEnvironmentsForLineno{flalign}%
\patchBothAmsMathEnvironmentsForLineno{alignat}%
\patchBothAmsMathEnvironmentsForLineno{gather}%
\patchBothAmsMathEnvironmentsForLineno{multline}%
\patchBothAmsMathEnvironmentsForLineno{eqnarray}%
}

\usepackage{hyperxmp}

\usepackage[pdftex,
            pdfauthor={\paperauthors},
            pdftitle={\paperasciititle},
            pdfkeywords={\paperkeywords},
            pdfcopyright={Copyright (C) \papercopyright},
            pdflicenseurl={\paperlicenceurl}]{hyperref}

\usepackage[colorinlistoftodos,textsize=scriptsize]{todonotes}

\usepackage[bottom,flushmargin,hang,multiple]{footmisc}

\usepackage[all]{hypcap} 

\input{lhcb-symbols-def} 

\usepackage{cite} 
\usepackage{mciteplus}

%% file: lhcb-symbols-def.tex
\usepackage{xspace} 
\usepackage{upgreek}

\def\lhcb   {\mbox{LHCb}\xspace}

\def\velo   {VELO\xspace}

\def\MagUp {\mbox{\em Mag\kern -0.05em Up}\xspace}

\ifthenelse{\boolean{uprightparticles}}%
{

 \def\Peta        {\ensuremath{\upeta}\xspace}

 \def\Pmu         {\ensuremath{\upmu}\xspace}                 
 \def\Pnu         {\ensuremath{\upnu}\xspace}                 
                  
 \def\Ppi         {\ensuremath{\uppi}\xspace}

 \def\PDelta      {\ensuremath{\Delta}\xspace}                 
 \def\PXi         {\ensuremath{\Xi}\xspace}                 
 \def\PLambda     {\ensuremath{\Lambda}\xspace}                 
 \def\PSigma      {\ensuremath{\Sigma}\xspace}                 
 \def\POmega      {\ensuremath{\Omega}\xspace}                 
 \def\PUpsilon    {\ensuremath{\Upsilon}\xspace}

 \def\PB      {\ensuremath{\mathrm{B}}\xspace}                 
                  
 \def\PD      {\ensuremath{\mathrm{D}}\xspace}

 \def\PK      {\ensuremath{\mathrm{K}}\xspace}

 \def\Pb      {\ensuremath{\mathrm{b}}\xspace}                 
 \def\Pc      {\ensuremath{\mathrm{c}}\xspace}                 
 \def\Pd      {\ensuremath{\mathrm{d}}\xspace}                 
 \def\Pe      {\ensuremath{\mathrm{e}}\xspace}

 \def\Pi      {\ensuremath{\mathrm{i}}\xspace}

 \def\Pp      {\ensuremath{\mathrm{p}}\xspace}

 \def\Ps      {\ensuremath{\mathrm{s}}\xspace}                 
                  
 \def\Pu      {\ensuremath{\mathrm{u}}\xspace}

 \def\thebaroffset{0.0em}
}
{

 \def\Peta        {\ensuremath{\eta}\xspace}

 \def\Pmu         {\ensuremath{\mu}\xspace}                 
 \def\Pnu         {\ensuremath{\nu}\xspace}                 
                  
 \def\Ppi         {\ensuremath{\pi}\xspace}

 \mathchardef\PDelta="7101
 \mathchardef\PXi="7104
 \mathchardef\PLambda="7103
 \mathchardef\PSigma="7106
 \mathchardef\POmega="710A
 \mathchardef\PUpsilon="7107
                  
 \def\PB      {\ensuremath{B}\xspace}                 
                  
 \def\PD      {\ensuremath{D}\xspace}

 \def\PK      {\ensuremath{K}\xspace}

 \def\Pb      {\ensuremath{b}\xspace}                 
 \def\Pc      {\ensuremath{c}\xspace}                 
 \def\Pd      {\ensuremath{d}\xspace}                 
 \def\Pe      {\ensuremath{e}\xspace}

 \def\Pi      {\ensuremath{i}\xspace}

 \def\Pp      {\ensuremath{p}\xspace}

 \def\Ps      {\ensuremath{s}\xspace}                 
                  
 \def\Pu      {\ensuremath{u}\xspace}

 \def\thebaroffset{0.18em}
}
\newcommand{\offsetoverline}[2][\thebaroffset]{\kern #1\overline{\kern -#1 #2}}%

\makeatletter
\ifcase \@ptsize \relax
  \newcommand{\miniscule}{\@setfontsize\miniscule{4}{5}}
\or
  \newcommand{\miniscule}{\@setfontsize\miniscule{5}{6}}
\or
  \newcommand{\miniscule}{\@setfontsize\miniscule{5}{6}}
\fi
\makeatother

\DeclareRobustCommand{\optbar}[1]{\shortstack{{\miniscule (\rule[.5ex]{1.25em}{.18mm})}
  \\ [-.7ex] $#1$}}

\def\en         {{\ensuremath{\Pe^-}}\xspace}   
\def\ep         {{\ensuremath{\Pe^+}}\xspace}

\def\mup        {{\ensuremath{\Pmu^+}}\xspace}

\def\neu        {{\ensuremath{\Pnu}}\xspace}

\def\neue       {{\ensuremath{\neu_e}}\xspace}

\def\neum       {{\ensuremath{\neu_\mu}}\xspace}

\def\uquark    {{\ensuremath{\Pu}}\xspace}

\def\dquark    {{\ensuremath{\Pd}}\xspace}

\def\squark    {{\ensuremath{\Ps}}\xspace}

\def\cquark    {{\ensuremath{\Pc}}\xspace}

\def\bquark    {{\ensuremath{\Pb}}\xspace}

\def\pion   {{\ensuremath{\Ppi}}\xspace}
\def\piz    {{\ensuremath{\pion^0}}\xspace}
\def\pip    {{\ensuremath{\pion^+}}\xspace}
\def\pim    {{\ensuremath{\pion^-}}\xspace}

\def\kaon    {{\ensuremath{\PK}}\xspace}

\def\KorKbar {\kern \thebaroffset\optbar{\kern -\thebaroffset \PK}{}\xspace}

\def\Kp      {{\ensuremath{\kaon^+}}\xspace}
\def\Km      {{\ensuremath{\kaon^-}}\xspace}

\def\KS      {{\ensuremath{\kaon^0_{\mathrm{S}}}}\xspace}

\newcommand{\etaz}{\ensuremath{\Peta}\xspace}

\def\D       {{\ensuremath{\PD}}\xspace}

\def\DorDbar {\kern \thebaroffset\optbar{\kern -\thebaroffset \PD}\xspace}
\def\Dz      {{\ensuremath{\D^0}}\xspace}

\def\Dp      {{\ensuremath{\D^+}}\xspace}
\def\Dm      {{\ensuremath{\D^-}}\xspace}

\def\DpDm    {\ensuremath{\Dp {\kern -0.16em \Dm}}\xspace}

\def\Dsp     {{\ensuremath{\D^+_\squark}}\xspace}

\def\B       {{\ensuremath{\PB}}\xspace}

\def\BorBbar {\kern \thebaroffset\optbar{\kern -\thebaroffset \PB}\xspace}

\def\Bd      {{\ensuremath{\B^0}}\xspace}

\def\BdorBdbar {\kern \thebaroffset\optbar{\kern -\thebaroffset \Bd}\xspace}

\def\Bs      {{\ensuremath{\B^0_\squark}}\xspace}

\def\BsorBsbar {\kern \thebaroffset\optbar{\kern -\thebaroffset \Bs}\xspace}

\def\Y#1S{\ensuremath{\PUpsilon{(#1S)}}\xspace}

\def\proton      {{\ensuremath{\Pp}}\xspace}

\def\Lz          {{\ensuremath{\PLambda}}\xspace}

\def\LorLbar     {\kern \thebaroffset\optbar{\kern -\thebaroffset \PLambda}\xspace}

\def\Lc          {{\ensuremath{\Lz^+_\cquark}}\xspace}

\def\BF         {{\ensuremath{\mathcal{B}}}\xspace}

\newcommand{\decay}[2]{\ensuremath{#1\!\to #2}\xspace} 

\def\to                 {\ensuremath{\rightarrow}\xspace}

\def\order   {{\ensuremath{\mathcal{O}}}\xspace}

\def\CP                {{\ensuremath{C\!P}}\xspace}

\def\Vcd  {{\ensuremath{V_{\cquark\dquark}^{\phantom{\ast}}}}\xspace}

\def\Vcs  {{\ensuremath{V_{\cquark\squark}^{\phantom{\ast}}}}\xspace}

\def\Vuds  {{\ensuremath{V_{\uquark\dquark}^\ast}}\xspace}

\def\Vuss  {{\ensuremath{V_{\uquark\squark}^\ast}}\xspace}

\def\AT#1     {\ensuremath{A_{\mathrm{T}}^{#1}}\xspace}           

\def\C#1      {\ensuremath{\mathcal{C}_{#1}}\xspace}                       
\def\Cp#1     {\ensuremath{\mathcal{C}_{#1}^{'}}\xspace}                    
\def\Ceff#1   {\ensuremath{\mathcal{C}_{#1}^{\mathrm{(eff)}}}\xspace}        
\def\Cpeff#1  {\ensuremath{\mathcal{C}_{#1}^{'\mathrm{(eff)}}}\xspace}       
\def\Ope#1    {\ensuremath{\mathcal{O}_{#1}}\xspace}                       
\def\Opep#1   {\ensuremath{\mathcal{O}_{#1}^{'}}\xspace}                    

\newcommand{\nospaceunit}[1]{\ensuremath{\text{#1}}}       
\newcommand{\aunit}[1]{\ensuremath{\text{\,#1}}}       

\newcommand{\tev}{\aunit{Te\kern -0.1em V}\xspace}
\newcommand{\gev}{\aunit{Ge\kern -0.1em V}\xspace}
\newcommand{\mev}{\aunit{Me\kern -0.1em V}\xspace}
\newcommand{\kev}{\aunit{ke\kern -0.1em V}\xspace}
\newcommand{\ev}{\aunit{e\kern -0.1em V}\xspace}
 
\newcommand{\mevc}{\ensuremath{\aunit{Me\kern -0.1em V\!/}c}\xspace}
\newcommand{\gevc}{\ensuremath{\aunit{Ge\kern -0.1em V\!/}c}\xspace}
\newcommand{\mevcc}{\ensuremath{\aunit{Me\kern -0.1em V\!/}c^2}\xspace}
\newcommand{\gevcc}{\ensuremath{\aunit{Ge\kern -0.1em V\!/}c^2}\xspace}

\def\mum  {\ensuremath{\,\upmu\nospaceunit{m}}\xspace}

\def\fb   {\ensuremath{\aunit{fb}}\xspace}
\def\invfb   {\ensuremath{\fb^{-1}}\xspace}

\def\ps   {\ensuremath{\aunit{ps}}\xspace}

\def\order{{\ensuremath{\mathcal{O}}}\xspace}

\def\gsim{{~\raise.15em\hbox{$>$}\kern-.85em
          \lower.35em\hbox{$\sim$}~}\xspace}
\def\lsim{{~\raise.15em\hbox{$<$}\kern-.85em
          \lower.35em\hbox{$\sim$}~}\xspace}

\def\sPlot{\mbox{\em sPlot}\xspace}

\def\pt         {\ensuremath{p_{\mathrm{T}}}\xspace}

\def\ptot       {\ensuremath{p}\xspace}

\def\mrad{\aunit{mrad}\xspace}

\def\evtgen     {\mbox{\textsc{EvtGen}}\xspace}

\def\geant      {\mbox{\textsc{Geant4}}\xspace}

\def\photos     {\mbox{\textsc{Photos}}\xspace}

\def\pythia     {\mbox{\textsc{Pythia}}\xspace}

\def\tell1  {TELL1\xspace}
\def\ukl1   {UKL1\xspace}



%% file: title-LHCb-PAPER.tex

\begin{titlepage}
\pagenumbering{roman}

\vspace*{-1.5cm}
\centerline{\large EUROPEAN ORGANIZATION FOR NUCLEAR RESEARCH (CERN)}
\vspace*{1.5cm}
\noindent
\begin{tabular*}{\linewidth}{lc@{\extracolsep{\fill}}r@{\extracolsep{0pt}}}
\ifthenelse{\boolean{pdflatex}}
{\vspace*{-1.5cm}\mbox{\!\!\!\includegraphics[width=.14\textwidth]{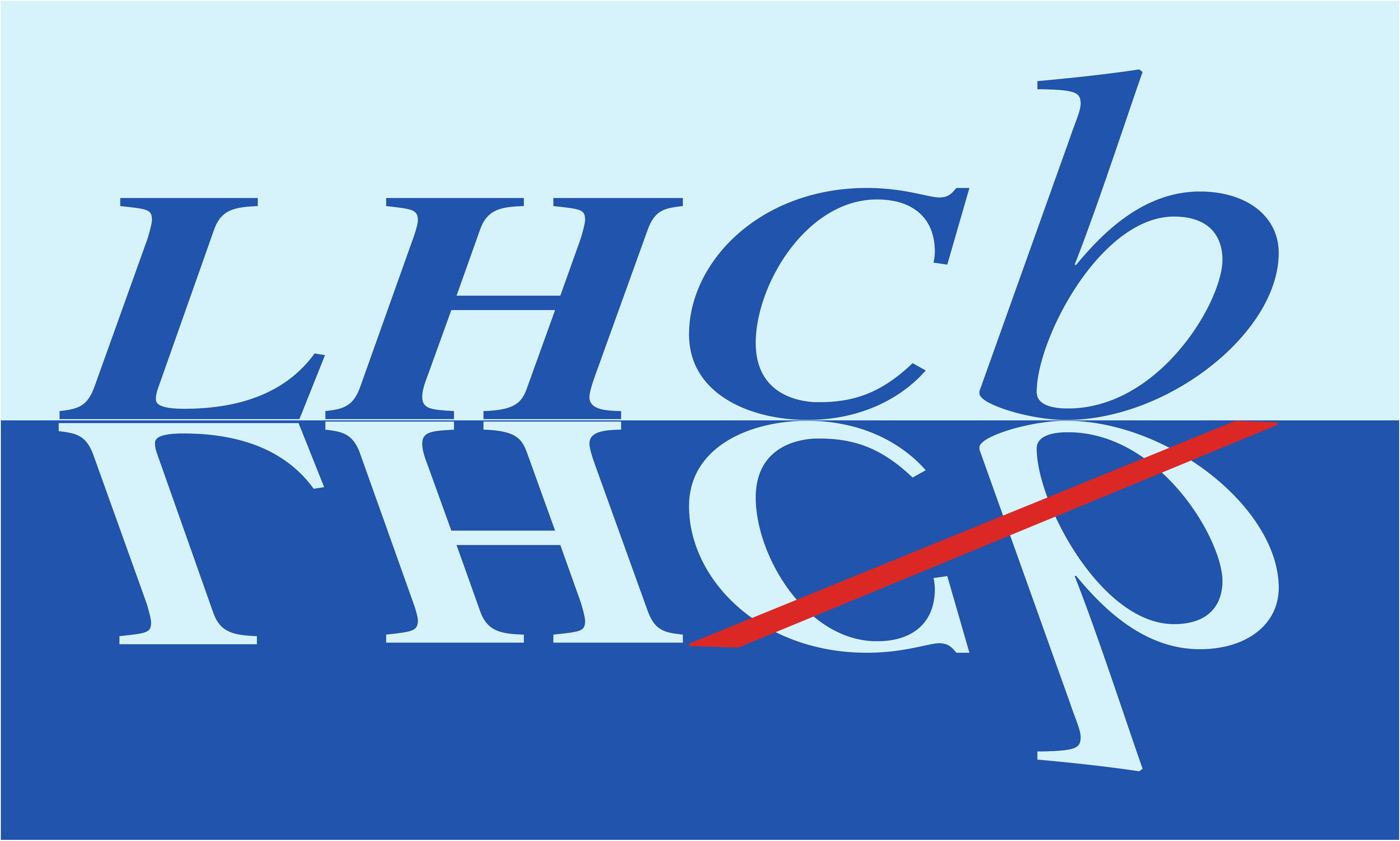}} & &}%
{\vspace*{-1.2cm}\mbox{\!\!\!\includegraphics[width=.12\textwidth]{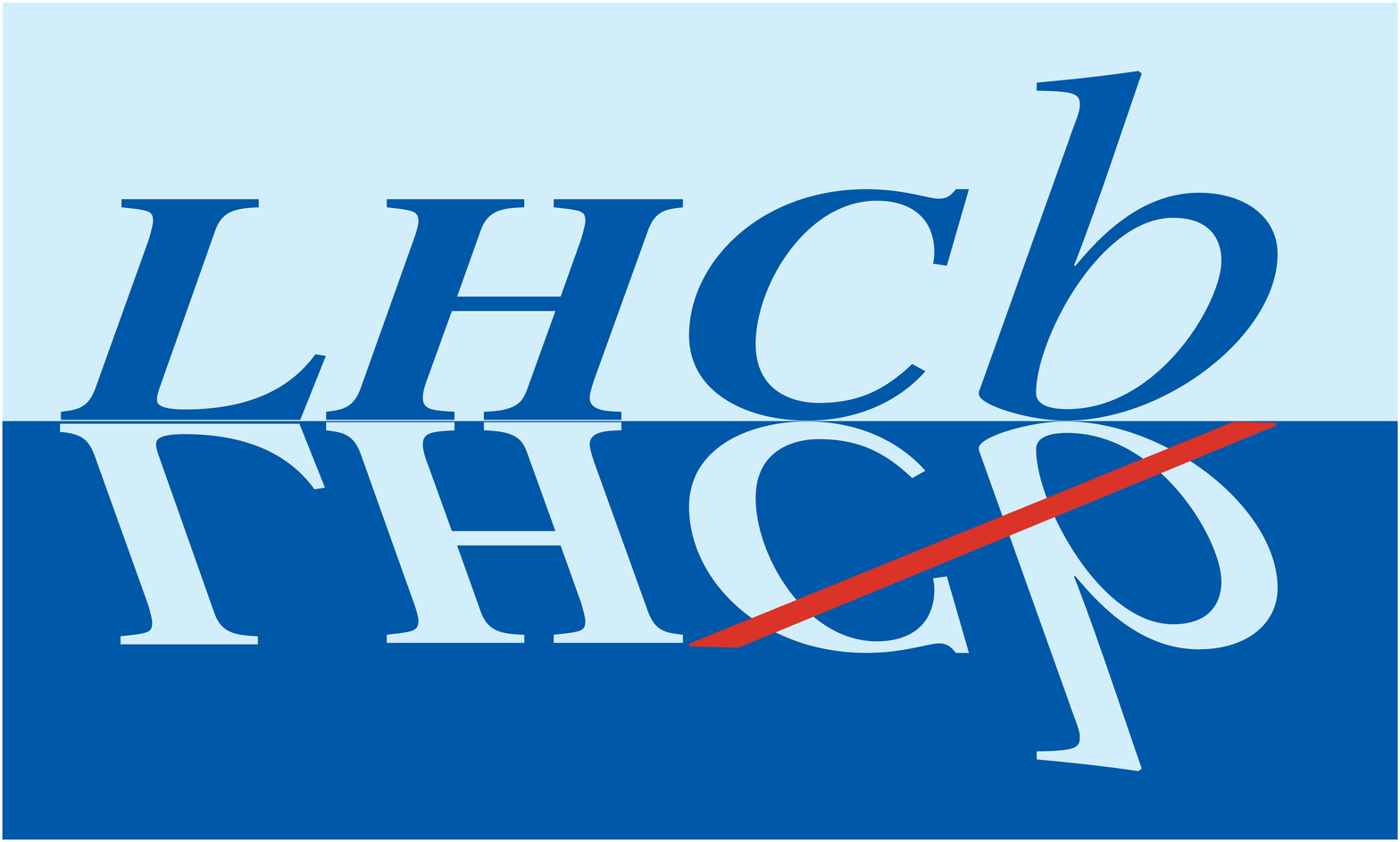}} & &}%
\\
 & & CERN-EP-2021-036 \\  
 & & LHCb-PAPER-2021-001 \\  
 & & June 14, 2021 \\ 
 & & \\
\end{tabular*}

\vspace*{2.8cm}

{\normalfont\bfseries\boldmath\huge
\begin{center}
  \papertitle 
\end{center}
}

\vspace*{1.3cm}

\begin{center}
\paperauthors\footnote{Authors are listed at the end of this paper.}
\end{center}

\vspace{\fill}

\begin{abstract}
\noindent Searches for \CP violation in the two-body decays $\decay{D_{(s)}^{+}}{h^{+}\piz}$ and $\decay{D_{(s)}^{+}}{h^{+}\eta}$ (where $h^{+}$ denotes a \pip or \Kp meson) are performed using \proton\proton collision data collected by the \lhcb experiment corresponding to either 9\invfb or 6\invfb of integrated luminosity. The \piz and \etaz mesons are reconstructed using the $\ep\en\gamma$ final state, which can proceed as  three-body decays \decay{\piz}{\ep\en\gamma} and \decay{\etaz}{\ep\en\gamma}, or via the two-body decays \decay{\piz}{\gamma\gamma} and \decay{\etaz}{\gamma\gamma} followed by a photon conversion. The measurements are made relative to the control modes $\decay{D_{(s)}^{+}}{\KS h^{+}}$ to cancel the production and detection asymmetries. The \CP asymmetries are measured to be
\begin{alignat*}{7}
    \mathcal{A}_{\CP}(\decay{\Dp}{\pip\piz}) 	&= (-&&1.3 &&\pm 0.9 &&\pm 0.6 &)\%, \\
    \mathcal{A}_{\CP}(\decay{\Dp}{\Kp\piz}) 	&= (-&&3.2 &&\pm 4.7 &&\pm 2.1 &)\%, \\
    \mathcal{A}_{\CP}(\decay{\Dp}{\pip\etaz})   &= (-&&0.2 &&\pm 0.8 &&\pm 0.4 &)\%, \\
    \mathcal{A}_{\CP}(\decay{\Dp}{\Kp\etaz}) 	&= (-&&6 &&\pm 10 &&\pm 4 &)\%, \\
    \mathcal{A}_{\CP}(\decay{\Dsp}{\Kp\piz}) 	&= (-&&0.8 &&\pm 3.9 &&\pm 1.2 &)\%, \\
    \mathcal{A}_{\CP}(\decay{\Dsp}{\pip\etaz})  &= (&&0.8 &&\pm 0.7 &&\pm 0.5 &)\%, \\
    \mathcal{A}_{\CP}(\decay{\Dsp}{\Kp\etaz})   &= (&&0.9 &&\pm 3.7 &&\pm 1.1 &)\%,
\end{alignat*}
where the first uncertainties are statistical and the second systematic. These results are consistent with no \CP violation and mostly constitute the most precise measurements of $\mathcal{A}_{\CP}$ in these decay modes to date.   
\end{abstract}

\vspace*{0.1cm}

\begin{center}
  Published in JHEP 06 (2021) 019
\end{center}

\vspace{\fill}

{\footnotesize 
\centerline{\copyright~\papercopyright. \href{\paperlicenceurl}{\paperlicence}.}}
\vspace*{2mm}

\end{titlepage}


\newpage
\setcounter{page}{2}
\mbox{~}
%
%
%
%

%% file: body.tex
\section{Introduction}
\label{sec:Introduction}

\begin{figure}[h]
    \centering
    \begin{subfigure}[t]{\linewidth}
        \centering
        \includegraphics[width=0.4\linewidth]{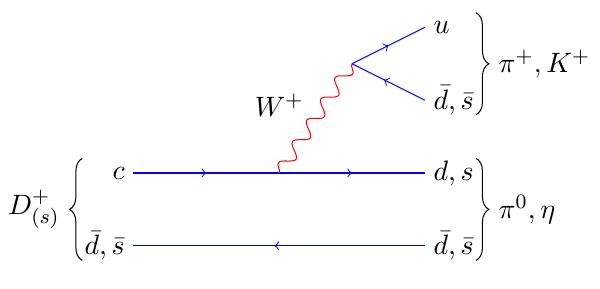}
        \includegraphics[width=0.4\linewidth]{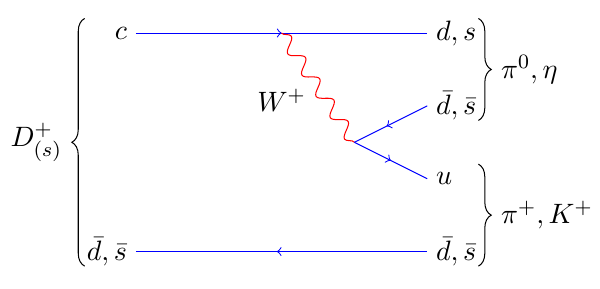}
    \end{subfigure}
    \begin{subfigure}[t]{\linewidth}
        \centering
        \includegraphics[width=0.4\linewidth]{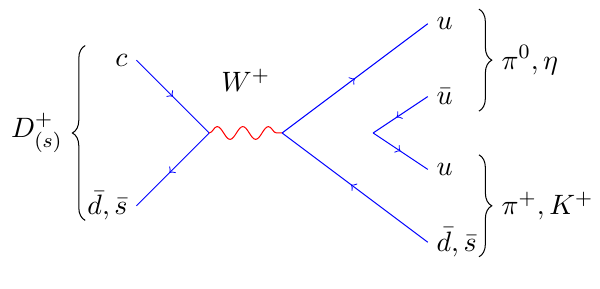}
        \includegraphics[width=0.4\linewidth]{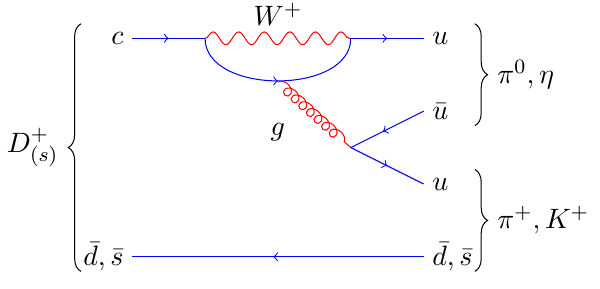}
    \end{subfigure}
    \caption{Processes that contribute to the studied decays at tree-level include (top left) colour-favoured, (top right) colour-suppressed and (bottom left) annihilation topology decays. Contributions can also be received at loop-level from processes such as (bottom right) penguin topology decays.  }
    \label{fig:feynman}
\end{figure}

The observation of Charge-Parity (\CP) violation in two-body decays of neutral \D mesons~\cite{LHCB-PAPER-2019-006} motivates searches for similar effects in charged \D meson decays. 
The two-body \mbox{$\decay{D_{(s)}^{+}}{h^{+}\piz}$} and $\decay{D_{(s)}^{+}}{h^{+}\eta}$ decays, where $h^{+}$ denotes a \pip or \Kp meson,\footnote{Inclusion of charge conjugated processes is implied throughout, except when discussing asymmetry definitions.} are mediated by Cabibbo favoured (CF), singly Cabibbo suppressed (SCS) or doubly Cabibbo suppressed (DCS) processes. The contributing decay topologies are shown in Fig.~\ref{fig:feynman}. The SCS modes \decay{\Dsp}{\Kp\piz}, \decay{\Dp}{\pip\etaz} and \decay{\Dsp}{\Kp\etaz} receive contributions from two different weak phases, proportional to the products of the CKM matrix elements $\Vcd\Vuds$ and $\Vcs\Vuss$, allowing \CP violation at tree-level. In the Standard Model (SM), the \CP asymmetries are expected to be of the order $10^{-4}$--$10^{-3}$~\cite{PhysRevD.85.034036,PhysRevD.86.036012,Pirtskhalava:2011va,PhysRevD.99.113001,Bhattacharya:2012ah,Bause:2020obd}. The CF mode \decay{\Dsp}{\pip\etaz} and the DCS modes \decay{\Dp}{\Kp\piz} and \decay{\Dp}{\Kp\etaz} receive contributions from only one weak phase at tree-level. The \decay{\Dsp}{\pip\piz} mode proceeds via an annihilation topology decay and is therefore highly suppressed.

The SCS \decay{\Dp}{\pip\piz} mode is of particular interest as the \CP asymmetry in the SM is expected to be zero as a result of isospin constraints~\cite{PhysRevD.86.036012,Pirtskhalava:2011va,PhysRevD.99.113001,Bhattacharya:2012ah}. The \CP asymmetries of the signal decays are defined to be
\begin{equation}
    \mathcal{A}_{\CP}(\decay{D_{(s)}^{+}}{h^{+}h^{0}}) \equiv \frac{\Gamma(\decay{D_{(s)}^{+}}{h^{+}h^{0}})- \Gamma(\decay{D_{(s)}^{-}}{h^{-}h^{0}})}{\Gamma(\decay{D_{(s)}^{+}}{h^{+}h^{0}})+\Gamma(\decay{D_{(s)}^{-}}{h^{-}h^{0}})},
\end{equation}
where $\Gamma$ is the partial decay rate and $h^{0}$ denotes either a \piz or an \etaz meson. 
A non-zero value of $\mathcal{A}_{\CP}(\decay{\Dp}{\pip\piz})$, coupled with a verification that the isospin sum rule 
\begin{equation}
R = \frac{\mathcal{A}_{\CP} (\decay{\Dz}{\pip\pim})}{1 + \frac{\tau_{\Dz}}{\BF_{+-}}\left(\frac{\BF_{00}}{\tau_{\Dz}} + \frac{2}{3}\frac{\BF_{+0}}{\tau_{\Dp}}\right) }
    + 
    \frac{\mathcal{A}_{\CP} (\decay{\Dz}{\piz\piz})}{1 + \frac{\tau_{\Dz}}{\BF_{00}}\left(\frac{\BF_{+-}}{\tau_{\Dz}} + \frac{2}{3}\frac{\BF_{+0}}{\tau_{\Dp}}\right) }
    - 
    \frac{\mathcal{A}_{\CP} (\decay{\Dp}{\pip\piz})}{1 + \frac{3}{2}\frac{\tau_{\Dp}}{\BF_{+0}}\left(\frac{\BF_{00}}{\tau_{\Dz}} + \frac{\BF_{+-}}{\tau_{\Dz}}\right) }
    \label{eq:sumrule}
\end{equation}
is consistent with zero, would be an indication of physics beyond the SM~\cite{Grossman:2012eb,Buccella:1992sg,Bause:2020obd,Babu:2017bjn}. Here, $\tau_{\Dp}$ and $\tau_{\Dz}$ represent the \Dp and \Dz lifetimes and $\BF_{+-}$, $\BF_{00}$ and $\BF_{+0}$ represent the branching fractions of $\decay{\Dz}{\pip\pim}$, $\decay{\Dz}{\piz\piz}$ and $\decay{\Dp}{\pip\piz}$ decays, respectively. 
A recent measurement from the Belle collaboration determined the \CP asymmetry to be
\mbox{$\mathcal{A}_{\CP}(\decay{\Dp}{\pip\piz}) =$ $(2.31\pm1.24\pm0.23)\%$}~\cite{Babu:2017bjn}, where the first uncertainty is statistical and the second is systematic, corresponding to a value of $R= (-2.2\pm2.7)\times 10^{-3}$.

In this article measurements of \CP asymmetries of seven \mbox{$\decay{D_{(s)}^{+}}{h^{+}\piz}$} and \mbox{$\decay{D_{(s)}^{+}}{h^{+}\eta}$} modes are performed, using samples corresponding to either 9\invfb or 6\invfb of integrated luminosity, respectively, collected by the \lhcb experiment in proton-proton (\proton\proton) collisions at the LHC.
The 6\invfb data set comprises data collected during 2015--2018 (Run~2) at a centre-of-mass energy of $13\tev$, whilst the 9\invfb data set additionally includes data collected during 2011--2012 (Run~1) at centre-of-mass energies of $7\tev$ and $8\tev$.     
The neutral \piz and \etaz mesons are reconstructed via decays to the $\ep\en\gamma$ final state. The reconstruction of electron and positron tracks, in addition to the charged hadron track from the $D^{+}_{(s)}$ meson decay, enables the determination of the displaced $D^{+}_{(s)}$ meson decay vertex and suppresses background from particles originating from the primary \proton\proton interaction. The signal receives contributions from the suppressed three-body Dalitz decays $\decay{\piz}{\ep\en\gamma}$ and $\decay{\eta}{\ep\en\gamma}$ with branching fractions $(1.174\pm0.035)\%$ and $(6.9\pm0.4)\times10^{-3}$, respectively~\cite{Zyla:2020zbs}, as well as the more common $\decay{\piz}{\gamma\gamma}$ and $\decay{\eta}{\gamma\gamma}$ decays with branching fractions $(98.823\pm0.034)\%$ and $(39.41\pm0.20)\%$~\cite{Zyla:2020zbs}, where one of the photons subsequently interacts with the detector material and is converted to an $\ep\en$ pair. 
Converted photons have been previously exploited at \lhcb~\cite{LHCb-PAPER-2013-028,Beaucourt:2016htk,LHCb-DP-2018-002,LHCb-PAPER-2021-003,Beteta:2020gui}, but this is the first measurement to use converted photons to reconstruct \piz and \etaz mesons.

Experimentally, the raw asymmetry of each signal mode is measured, which is defined to be 
\begin{equation}
    A_{\text{Raw}}(\decay{D_{(s)}^{+}}{h^{+}h^{0}}) \equiv \frac{N(\decay{D_{(s)}^{+}}{h^{+}h^{0}})- N(\decay{D_{(s)}^{-}}{h^{-}h^{0}})}{N(\decay{D_{(s)}^{+}}{h^{+}h^{0}})+N(\decay{D_{(s)}^{-}}{h^{-}h^{0}})},
\end{equation}
where $N$ is the signal yield. This can be approximated by
\begin{equation}
     A_{\text{Raw}}(\decay{D_{(s)}^{+}}{h^{+}h^{0}}) \approx \mathcal{A}_{\CP}(\decay{D_{(s)}^{+}}{h^{+}h^{0}}) + A_{\text{Prod}}(D_{(s)}^{+}) + A_{\text{Det}}(h^{+}),
\end{equation}
where $A_{\text{Prod}}(D_{(s)}^{+})$ and $A_{\text{Det}}(h^{+})$ represent the production and detection asymmetries of the corresponding hadrons.  
In order to cancel the production and detection asymmetries, the raw asymmetry of $\decay{D_{(s)}^{+}}{\KS h^{+}}$ control decays is subtracted, approximated by
\begin{equation}
     A_{\text{Raw}}(\decay{D_{(s)}^{+}}{\KS h^{+}}) \approx \mathcal{A}_{\CP}(\decay{D_{(s)}^{+}}{\KS h^{+}}) + A_{\text{Prod}}(D_{(s)}^{+}) + A_{\text{Det}}(h^{+}) + A_{\text{Mix}}(K^0),
\end{equation}
where the extra term $A_{\text{Mix}}(K^0)$ arises due to the \CP asymmetry induced by mixing and decay of the neutral \KS meson~\cite{Lipkin:1999qz}. 
As the nuisance asymmetries are known to be kinematically dependent, the $\decay{D_{(s)}^{+}}{\KS h^{+}}$ samples are weighted to match the kinematic distributions of the signal candidates to optimally reduce the impact of the production and detection asymmetries.
The \CP asymmetry for the signal modes can then be determined as
\begin{align}
\begin{split}
     \mathcal{A}_{\CP}(\decay{D_{(s)}^{+}}{h^{+}h^{0}}) ={} & 
     A_{\text{Raw}}(\decay{D_{(s)}^{+}}{h^{+}h^{0}}) - A_{\text{Raw}}^{\text{w}}(\decay{D_{(s)}^{+}}{\KS h^{+}})\\
     &+\mathcal{A}_{\CP}(\decay{D_{(s)}^{+}}{\KS h^{+}}) + A_{\text{Mix}}(K^0), 
\end{split}
\label{eq:result}
\end{align}
where $A_{\text{Raw}}^{\text{w}}$ represents the raw asymmetry determined from weighted samples, the values of $\mathcal{A}_{\CP}(\decay{D_{(s)}^{+}}{\KS h^{+}})$ are accounted for using external inputs with sub-percent precision~\cite{LHCb-PAPER-2019-002}, and $A_{\text{Mix}}(K^0)$ is calculated using a description of the detector material and the distribution of \KS decay times and momentum in the selected data, as detailed in Refs.~\cite{LHCb-PAPER-2019-002,LHCb-PAPER-2014-013}.

This article is structured as follows: the \lhcb experiment is described in Section~\ref{sec:Detector}; the requirements used to reconstruct the signal samples are given in Section~\ref{sec:Selection}; a description of the fits to the invariant mass distributions can be found in Section~\ref{sec:Fit_Model}; the treatment of the $\decay{D_{(s)}^{+}}{\KS h^{+}}$ control modes is given in Section~\ref{sec:Control_Modes}; the sources of systematic uncertainty are detailed in Section~\ref{sec:Systematic_Uncertainties}; and finally the results and conclusions are summarised in Section~\ref{sec:Results_and_conc}.
\section{Detector}
\label{sec:Detector}
The \lhcb detector~\cite{LHCb-DP-2008-001,LHCb-DP-2014-002} is a single-arm forward
spectrometer covering the \mbox{pseudorapidity} range between 2 and 5,
designed for the study of particles containing \bquark or \cquark
quarks. The detector includes a high-precision tracking system
consisting of a silicon-strip vertex detector surrounding the $pp$
interaction region (\velo), a large-area silicon-strip detector located
upstream of a dipole magnet with a bending power of about
$4{\mathrm{\,Tm}}$, and three stations of silicon-strip detectors and straw
drift tubes
placed downstream of the magnet.
The tracking system provides a measurement of the momentum, \ptot, of charged particles with
a relative uncertainty that varies from 0.5\% at low momentum to 1.0\% at 200\gevc.
The minimum distance of a track to a primary $pp$ collision vertex (PV), the impact parameter (IP), 
is measured with a resolution of $(15+29/\pt)\mum$,
where \pt is the component of the momentum transverse to the beam, in\,\gevc.
Different types of charged hadrons are distinguished using information
from two ring-imaging Cherenkov detectors. 
Photons, electrons and hadrons are identified by a calorimeter system consisting of
scintillating-pad and preshower detectors, an electromagnetic
and a hadronic calorimeter. Muons are identified by a
system composed of alternating layers of iron and multiwire
proportional chambers.
The online event selection is performed by a trigger, 
which consists of a hardware stage, based on information from the calorimeter and muon
systems, followed by a software stage, which applies a full event
reconstruction.

Simulation is required to determine the invariant-mass distributions of the signal decays, develop the selection and constrain the yields of background from other particles misidentified as the signal-decay products.
In the simulation, $pp$ collisions are generated using
\pythia~\cite{Sjostrand:2007gs} 
with a specific \lhcb configuration~\cite{LHCb-PROC-2010-056}.
Decays of unstable particles
are described by \evtgen~\cite{Lange:2001uf}, in which final-state
radiation is generated using \photos~\cite{davidson2015photos}.
The interaction of the generated particles with the detector, and its response,
are implemented using the \geant
toolkit~\cite{Allison:2006ve, *Agostinelli:2002hh} as described in
Ref.~\cite{LHCb-PROC-2011-006}. 
The underlying $pp$ interaction is reused multiple times, with an independently generated signal decay for each~\cite{LHCb-DP-2018-004}.
\section{Event selection}
\label{sec:Selection}

To reconstruct the $\D_{(s)}^+$ meson candidate a well-identified kaon or pion  track is combined with a neutral meson to form a secondary decay vertex displaced from any PV. The neutral \piz and \etaz candidates are formed from two oppositely charged electron tracks that are combined with a photon candidate to create a neutral-meson decay vertex. A bremsstrahlung-recovery algorithm associates additional deposits from soft photons to those produced by the electrons in the electromagnetic calorimeter. To improve the resolution, the electron tracks must include a track segment within the \velo.     

At the hardware trigger level, candidates are selected by either directly identifying high transverse-momentum deposits from the signal in the electromagnetic or hadronic calorimeters, or by independently identifying another energetic particle produced in the \proton\proton collision. Inclusive multivariate (MVA) software triggers ensure the presence of well-reconstructed tracks that are inconsistent with originating from any PV. 
A second high-level software trigger performs a full event reconstruction to form the $D^{+}_{(s)}$ candidates.  
In Run 1, no dedicated exclusive triggers for the signal modes were implemented, but small samples of $\decay{D_{(s)}^{+}}{h^{+}\piz}$ candidates are reconstructable as a result of the overlap with existing exclusive two- and three-body \D-meson-decay triggers. No attempt is made to reconstruct $\decay{D_{(s)}^{+}}{h^{+}\etaz}$ candidates using the Run 1 data set. 
In Run 2, dedicated exclusive software triggers were added to form both $\decay{D_{(s)}^{+}}{h^{+}\piz}$ and $\decay{D_{(s)}^{+}}{h^{+}\etaz}$ signal candidates. 
These require the presence of a photon and three well-reconstructed tracks, inconsistent with originating from any PV. 
The invariant masses of the \piz (\etaz) meson candidates are required to be in the range $70 < m(\ep\en\gamma) < 210 \mevcc$ ($450 < m(\ep\en\gamma) < 650 \mevcc$) with $\pt>200\mevc$ ($500\mevc$). 
The $D^{+}_{(s)}$ candidate is required to have a transverse momentum $\pt>3000\mevc$ and a good quality vertex with an associated p-value of greater than 0.0018, created by first combining the $\ep\en$ candidates to form the photon conversion or $h^{0}$ Dalitz decay vertex, which is then further combined with a photon and charged hadron to create the $D^{+}_{(s)}$ decay vertex.

Offline, the $D^{+}_{(s)}$ candidate selection is refined by requiring that the momentum of the tracks is in the range $3<\ptot<100\gevc$ and their pseudorapidity is between $1.5$ and $5.0$.  
The $D^{+}_{(s)}$ candidates are required to have a mass in the range $1600<m(D^{+}_{(s)})<2200 \mevcc$, be consistent with originating at a primary interaction and have a proper decay time of $t > 0.15\ps$ ($0.25\ps$) for $\decay{D_{(s)}^{+}}{h^{+}\piz}$ ($\decay{D_{(s)}^{+}}{h^{+}\eta}$) candidates. Additionally, the angle between the momentum direction and the vector joining the PV and $D^{+}_{(s)}$ decay vertex, referred to as the direction angle, must be smaller than 10\mrad. 

Fiducial requirements are placed on the charged-hadron tracks to remove regions of large detection asymmetries, for example regions where a track of one charge would be bent out of the acceptance by the magnetic field whilst the opposite charge would be detected; the same criteria are used as in the previous measurements of the control modes~\cite{LHCb-PAPER-2019-002}. 

Particle identification (PID) requirements are applied using MVA-based PID variables for the charged particles and the photon to reduce the amount of combinatorial and misidentification background~\cite{LHCb-DP-2018-001,LHCb-PUB-2016-021}. 
Loose PID requirements are applied to the pion and electron tracks. Tighter requirements are applied to kaon candidates to reduce the rate of \decay{\pip}{\Kp} misidentification from the more abundant pion modes into the suppressed kaon modes. 
When reconstructing \piz mesons, a loose requirement is placed on an MVA-based photon-quality variable~\cite{CalvoGomez:2042173}, whilst for \etaz mesons, a tighter condition is required to reduce the level of combinatorial background. 
Requirements are placed on electron bremsstrahlung PID variables that match the bremsstrahlung calorimeter deposit to the electron track before passing through the magnetic field to ensure that the correct photon deposits are recovered. 
Decays with a total of either zero or one bremsstrahlung photon per $\ep\en$ pair are used in this analysis. For $\decay{\Dp}{h^{+}\piz}$ ($\decay{\Dp}{h^{+}\etaz}$) decays this corresponds to 62\% and 38\% (31\% and 48\%) of the reconstructed candidates, respectively.   
Decays with two or more bremsstrahlung photons per $\ep\en$ pair are removed as they result in a poor $D_{(s)}^{+}$ invariant mass resolution and high background level.

The same offline selection requirements are used for candidates selected with different numbers of bremsstrahlung photons, and also between candidates decaying via photon conversions or three-body $\decay{h^{0}}{\ep\en\gamma}$ decays. The requirements give a reasonable compromise between the efficiency of each type of decay with efficiencies of the order $\order(10^{-6})$ in Run~1 and  $\order(10^{-5})$ in Run~2.

After the full selection has been applied, approximately 3\% (2\%) of events are found to have multiple $\decay{D_{(s)}^{+}}{h^{+}\piz}$ ($\decay{D_{(s)}^{+}}{h^{+}\etaz}$) candidates predominately due to combinations with alternative photon candidates, of which all are retained. The signal decays are found to be dominated by $\decay{\piz}{\gamma\gamma}$ and $\decay{\eta}{\gamma\gamma}$ decays followed by a photon conversion, rather than the three-body Dalitz decays $\decay{\piz}{\ep\en\gamma}$ and $\decay{\eta}{\ep\en\gamma}$, with approximately 86\% of the candidates resulting from photon conversions.

\section{Signal modes and fit model}
\label{sec:Fit_Model}
\begin{figure}[tbp]
    \centering
    \begin{subfigure}[t]{\linewidth}
        \centering
        \includegraphics[width=0.49\linewidth]{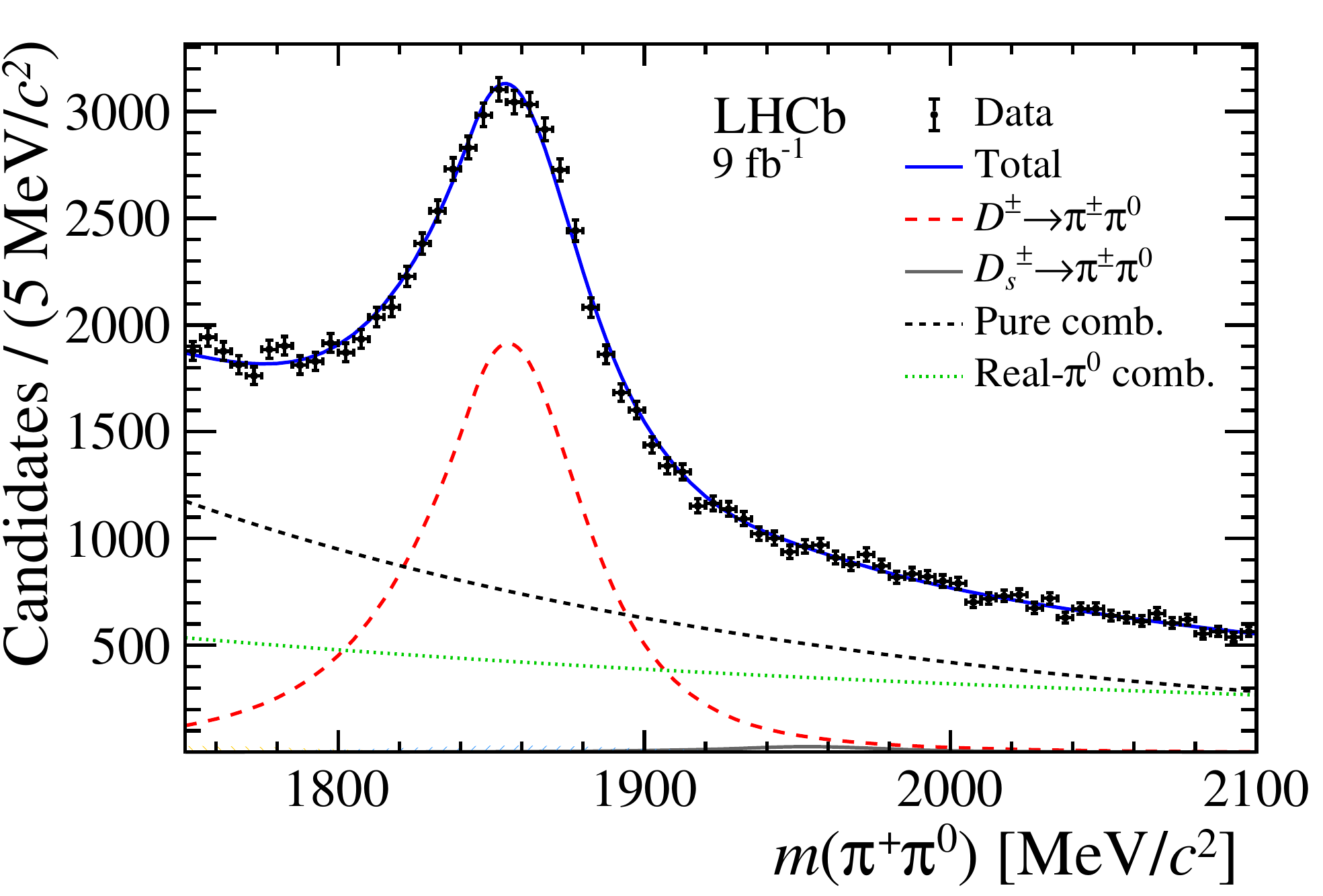}
        \includegraphics[width=0.49\linewidth]{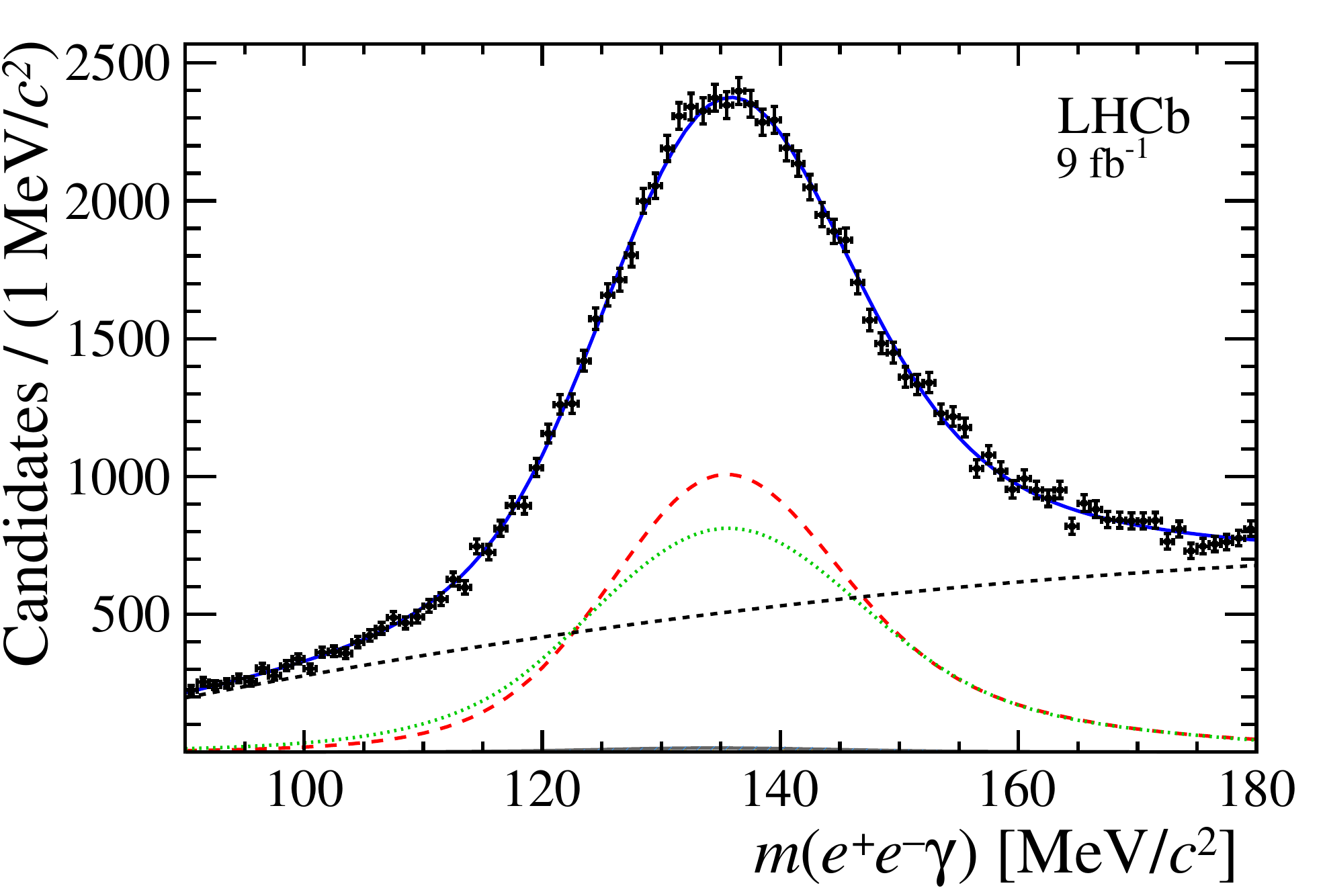}
    \end{subfigure}
    \begin{subfigure}[t]{\linewidth}
        \centering
        \includegraphics[width=0.49\linewidth]{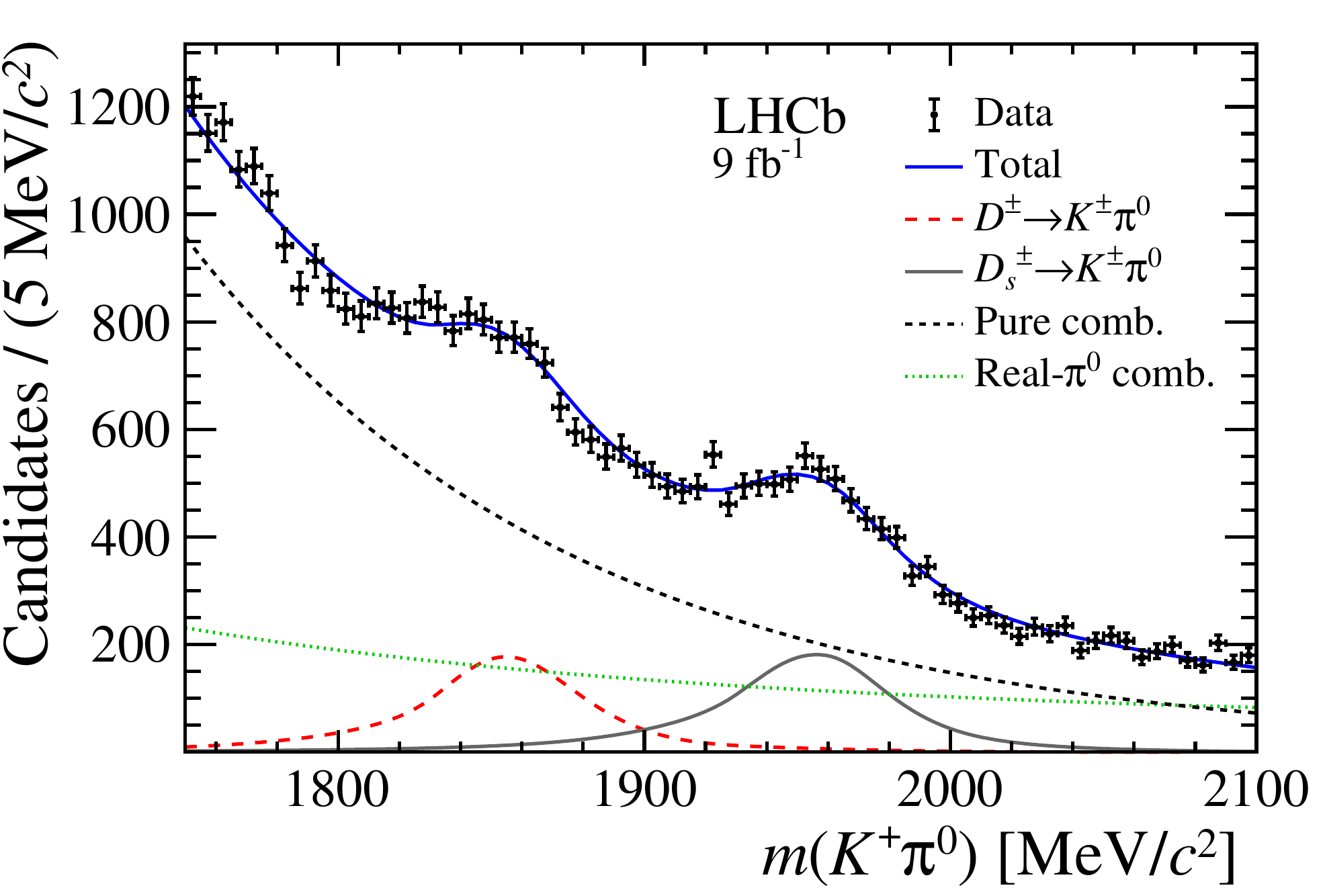}
        \includegraphics[width=0.49\linewidth]{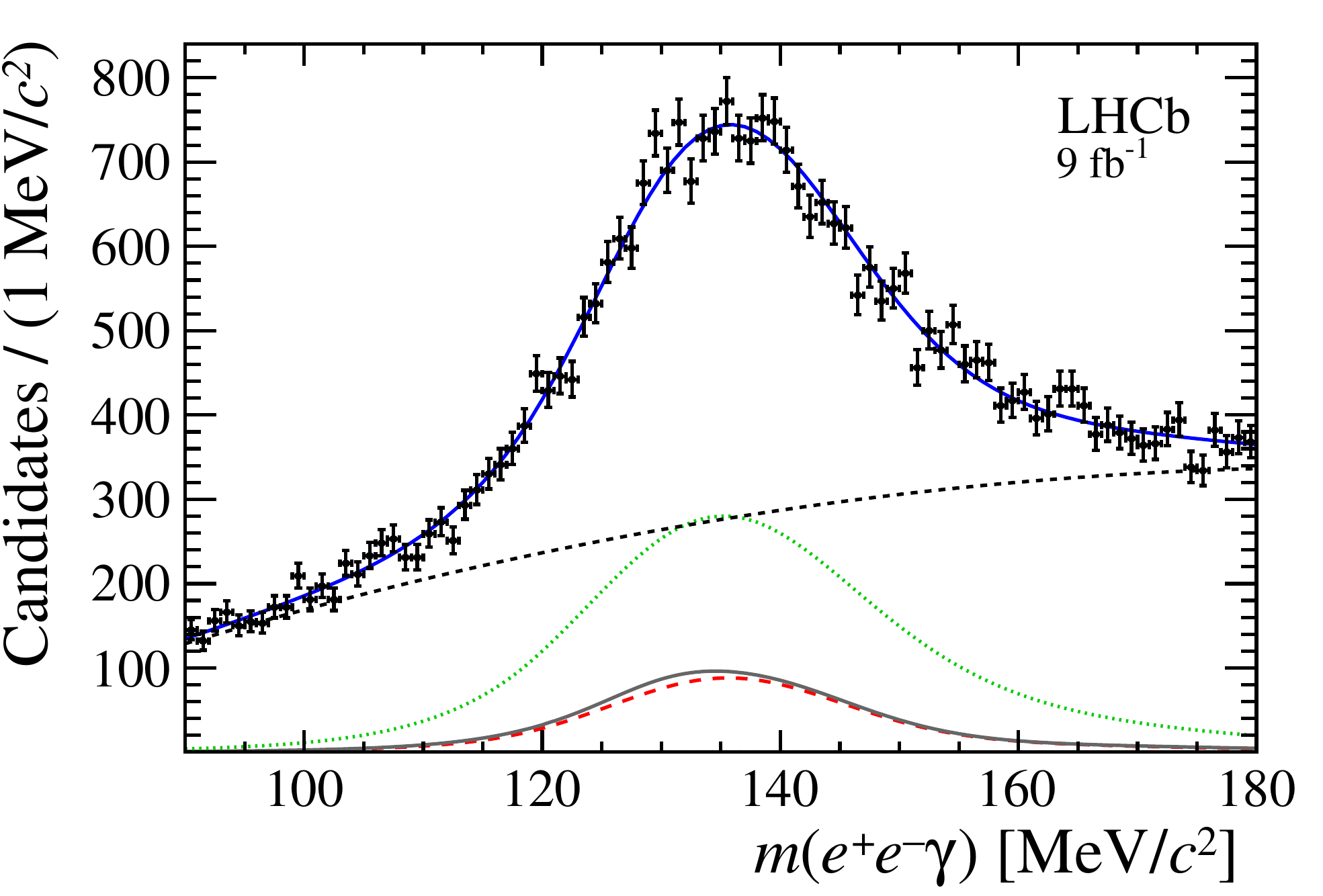}
    \end{subfigure}
    \caption{Distribution of the (left) $m(h^{+}\piz)$  and (right) $m(\ep\en\gamma)$  mass for (top) $\decay{D_{(s)}^{+}}{\pip\piz}$  and (bottom) $\decay{D_{(s)}^{+}}{\Kp\piz}$ candidates, summed over all categories of the simultaneous fit. 
    Projections of the total fit result and individual fit components are overlaid. This includes $\decay{\Dp}{h^{+}\piz}$ decays in dashed red, $\decay{\Dsp}{h^{+}\piz}$ decays in solid grey, pure combinatorial decays in dashed black and real-\piz combinatorial background in dotted green. The misidentification background is too small to be seen in these distributions. }
    \label{fig:results_pi0}
\end{figure}

\begin{figure}[tbp]
    \centering
    \begin{subfigure}[t]{\linewidth}
        \centering
        \includegraphics[width=0.49\linewidth]{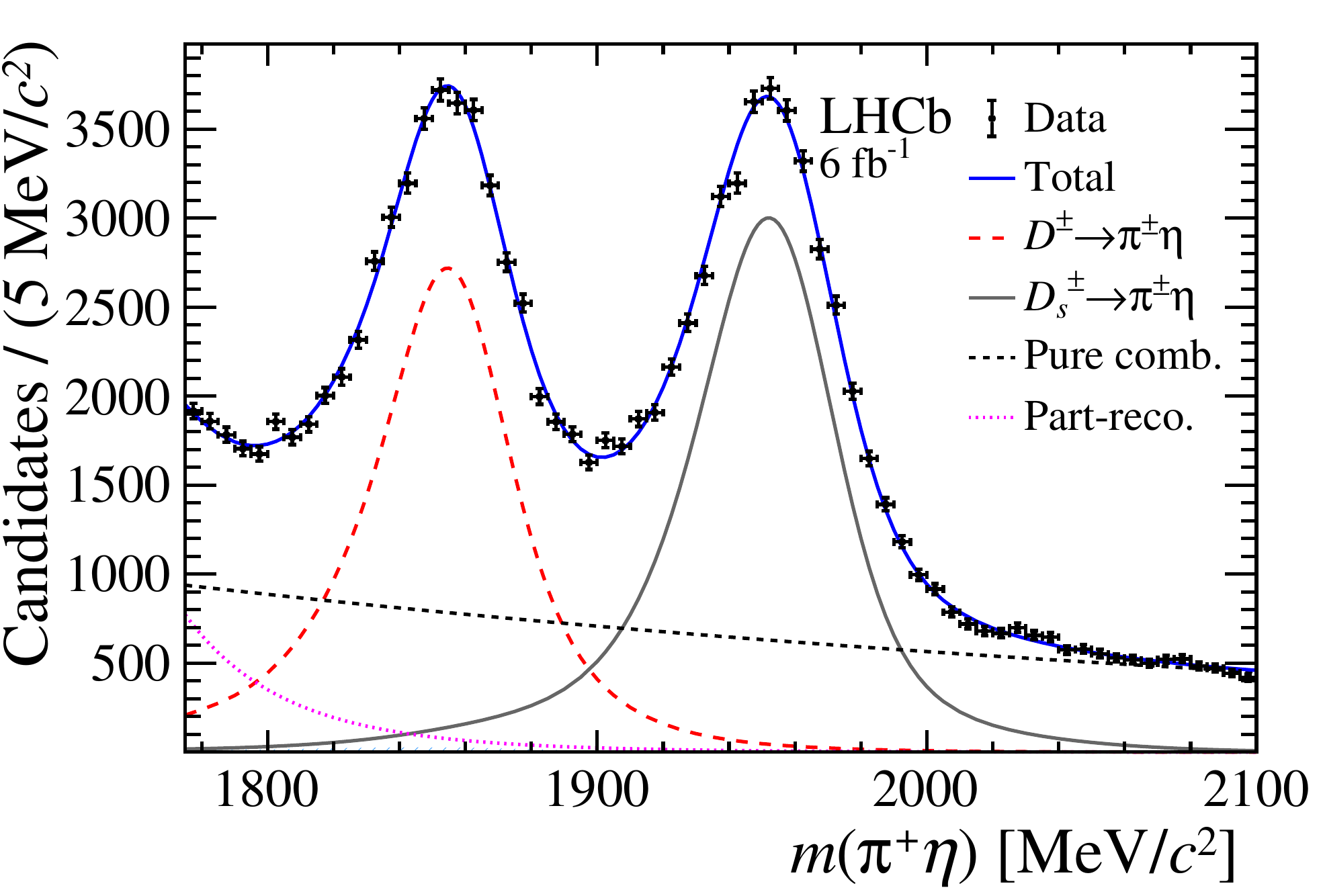}
        \includegraphics[width=0.49\linewidth]{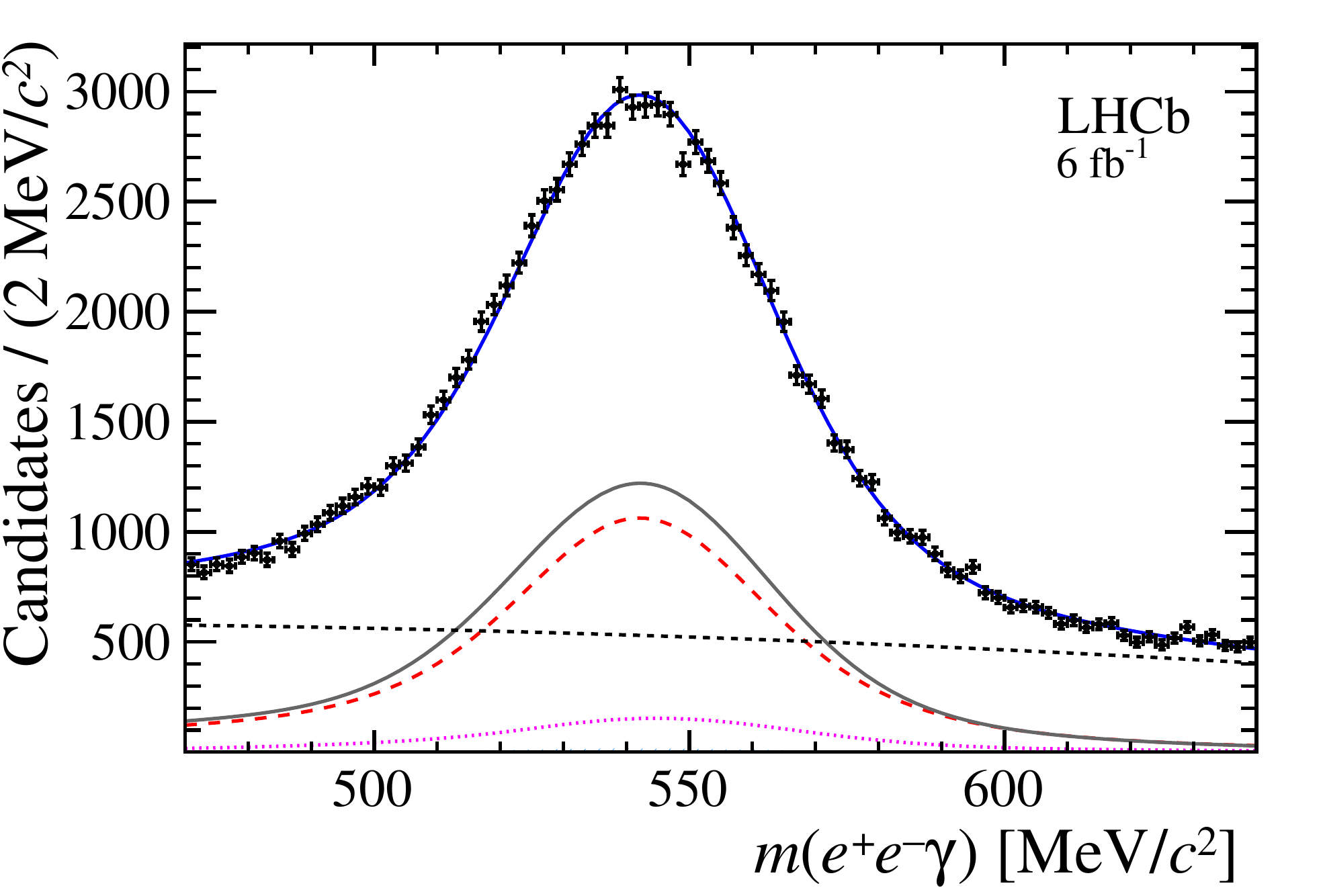}
    \end{subfigure}
    \begin{subfigure}[t]{\linewidth}
        \centering
        \includegraphics[width=0.49\linewidth]{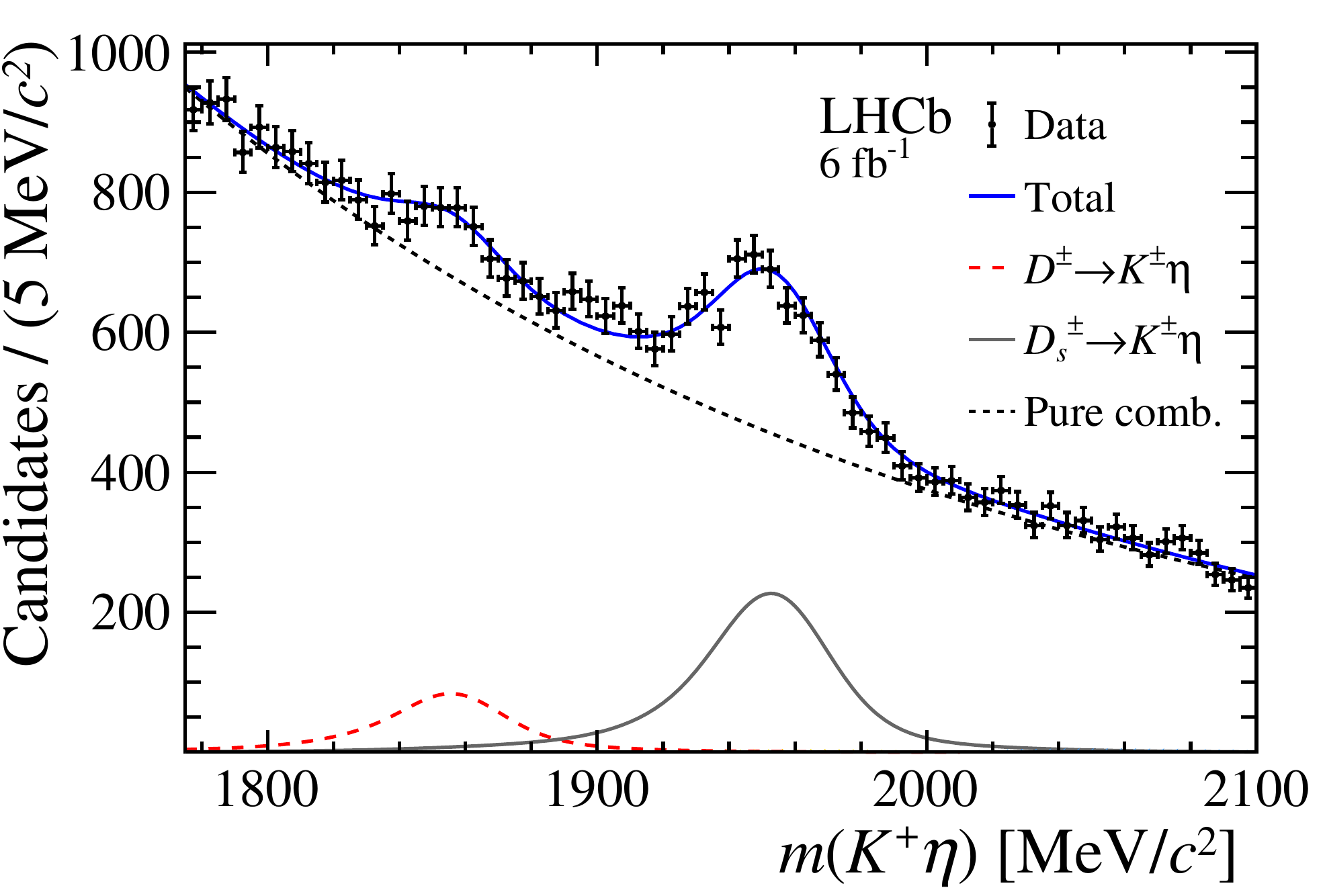}
        \includegraphics[width=0.49\linewidth]{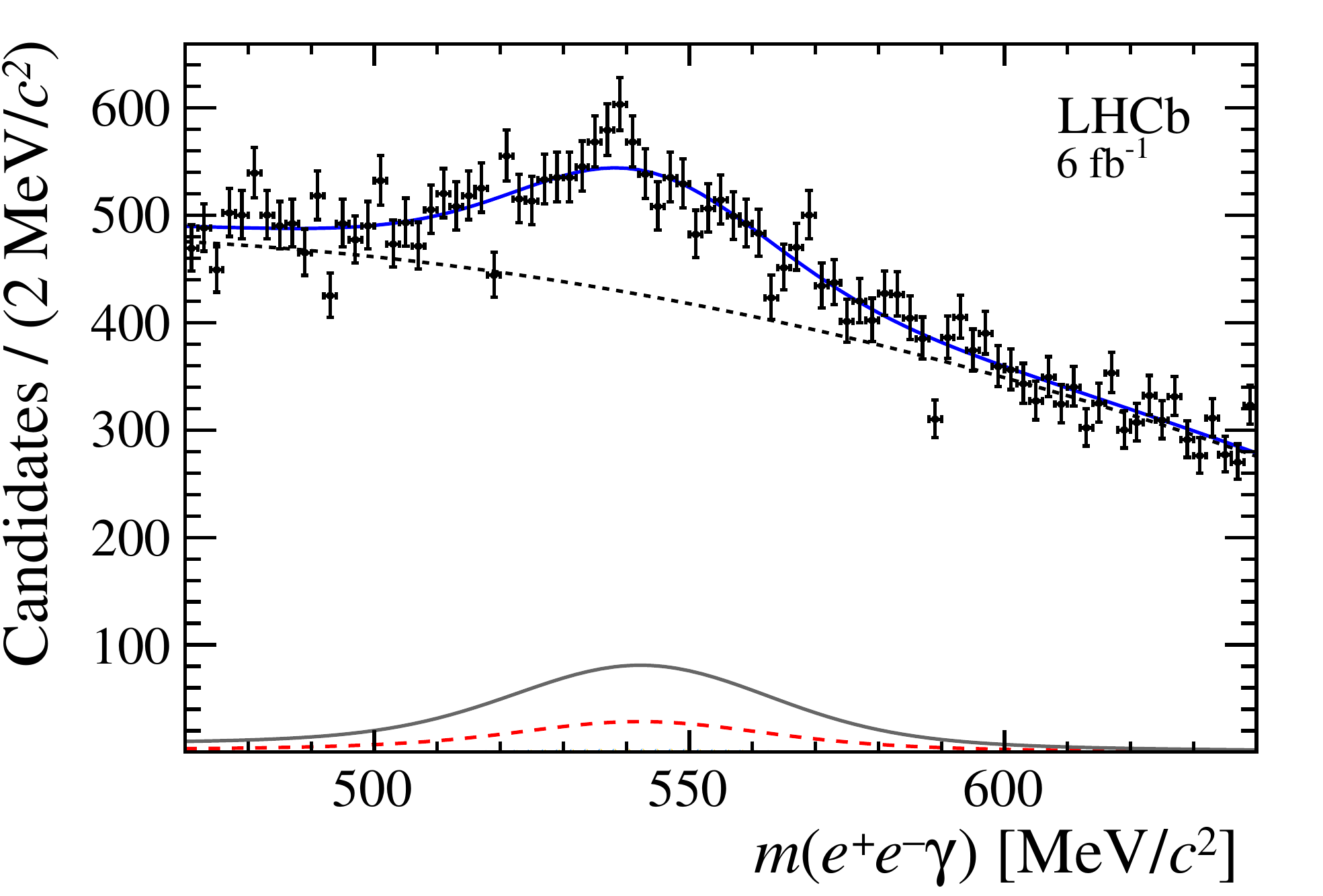}
    \end{subfigure}
    \caption{Distribution of the (left) $m(h^{+}\etaz)$ and (right) $m(\ep\en\gamma)$ mass  for (top) $\decay{D_{(s)}^{+}}{\pip\etaz}$ and (bottom) $\decay{D_{(s)}^{+}}{\Kp\etaz}$  candidates, summed over all categories of the simultaneous fit. 
    Projections of the total fit result and individual fit components are overlaid. This includes $\decay{\Dp}{h^{+}\etaz}$ decays in dashed red, $\decay{\Dsp}{h^{+}\etaz}$ decays in solid grey, pure combinatorial decays in dashed black and partially reconstructed background in dotted magenta. The misidentification background is too small to be seen in these distributions. }
    \label{fig:results_eta}
\end{figure}

The raw asymmetries of the signal modes are measured using two-dimensional extended simultaneous unbinned maximum-likelihood fits to the invariant mass $m(\ep\en\gamma)$ and the invariant mass difference $m(h^{+}h^{0}) \equiv m(h^+\ep\en\gamma) - m(\ep\en\gamma) + M(h^0)_{\text{PDG}}$, where $M(h^0)_{\text{PDG}}$ corresponds to the known \piz and \etaz masses~\cite{Zyla:2020zbs}. The quantity $m(h^{+}h^{0})$ is constructed to reduce the correlations between the two dimensions, and is referred to as the $D_{(s)}^{+}$ candidate mass henceforth. The $m(h^{+}h^{0})$ and $m(\ep\en\gamma)$ mass distributions are shown for $\decay{D_{(s)}^{+}}{h^{+}\piz}$ and $\decay{D_{(s)}^{+}}{h^{+}\etaz}$ candidates in Figs.~\ref{fig:results_pi0} and~\ref{fig:results_eta}.
The fits are performed for $\decay{D_{(s)}^{+}}{h^{+}\piz}$ candidates in the ranges \mbox{$1750 < m(h^{+}h^{0}) < 2100\mevcc$} and \mbox{$90 < m(\ep\en\gamma) < 180 \mevcc$}, and for $\decay{D_{(s)}^{+}}{h^{+}\etaz}$ candidates in the ranges \mbox{$1775 < m(h^{+}h^{0}) < 2100\mevcc$} and \mbox{$470 < m(\ep\en\gamma) < 640 \mevcc$}.

%
The fits are performed simultaneously on candidates  in categories that depend on the running period, the presence of bremsstrahlung photons, charged-hadron type (pion or kaon) and the candidate charge. All $\decay{D_{(s)}^{+}}{h^{+}\etaz}$ candidates were collected during Run 2. The $\decay{D_{(s)}^{+}}{h^{+}\piz}$ candidates are split into three running period categories, 2011, 2012 and Run 2, where the centre-of-mass energies were 7, 8, and 13\tev, respectively. Candidates with either zero or one bremsstrahlung photon per $\ep\en$ pair are split into two categories as they have different mass resolutions. The fits are performed on candidates with \pip and \Kp mesons simultaneously to allow the signal yields in either category to determine the misidentification-background yields in the corresponding category.

Two-dimensional probability density functions (PDFs) are used to model different contributions within the mass windows. These contributions can be categorised as signal decays, misidentification background, partially reconstructed low-mass  background and combinatorial background. 
The sum of positively- and negatively-charged candidate yields and raw asymmetry of all signal and background components are free to vary in the fits. A component for $\decay{\Dsp}{\pip\piz}$ signal is included in the fit, but due to the insignificant yield no corresponding raw asymmetry is measured. The PDFs are assumed to be the same for positively and negatively charged candidates, but otherwise allowed to differ for the other categories of the simultaneous fit. In the fit to $\decay{D^{+}_{(s)}}{h^{+}\piz}$ candidates the same raw asymmetries are shared between different running periods.

The signal modes are modelled by the sum of a two-dimensional Gaussian function and two two-dimensional Crystal Ball functions~\cite{Skwarnicki:1986xj}. The shape parameters and fraction of each function are determined from fits to simulated decays passing the full selection. To account for residual correlations between $m(h^{+}h^{0})$ and $m(\ep\en\gamma)$ resulting in part from radiative tails, the mean $h^0$ ($D^{+}_{(s)}$) mass is allowed to vary quadratically as a function of the $D^{+}_{(s)}$ ($h^0$) mass in the fits to $\decay{D_{(s)}^{+}}{h^{+}\piz}$ ($\decay{D_{(s)}^{+}}{h^{+}\etaz}$) candidates. 
When performing fits to data, freely varying scaling factors are applied to the widths of the PDFs, and freely varying offsets are added to the mean positions and quadratic correlation coefficients to account for differences between data and simulation. Different parameters are introduced for each running period and bremsstrahlung category. 
When determining PDFs from simulated decays, the candidates are weighted to account for the PID requirements using input from calibration samples~\cite{LHCb-PUB-2016-021}. 

The fit model accounts for misidentified signal decays, where a \pip track has been incorrectly assigned the \Kp mass hypothesis, or vice versa, using the same two-dimensional parameterisation as the signal shapes. The PDF parameters are determined from fits to the corresponding simulated signal decays passing the full selection for the charged hadron with the wrong mass hypothesis, including weights to account for the misidentification probabilities. When performing the fits to data, the yield of the misidentification background is constrained to the yield of signal in the other charged-hadron category multiplied by the relevant ratio of efficiencies determined from simulated decays and PID calibration samples. The yields of misidentification background contributions are below approximately 3\% of the corresponding signal yields.

Combinatorial background resulting from random combinations of tracks and photons is modelled with an exponential function in the $m(h^{+}h^{0})$ dimension and a second-order Chebychev polynomial function in the $m(\ep\en\gamma)$ dimension. The exponential coefficient and Chebychev polynomial coefficients freely vary in the fit. 
In the fit to $\decay{D_{(s)}^{+}}{h^{+}\piz}$ candidates, it is found necessary to include a combinatorial component comprising a real \piz meson combined with an unrelated track. The PDF is constructed from a peaking distribution in the $m(\ep\en\gamma)$ dimension and an exponential function in the $m(h^{+}h^{0})$  dimension. The peaking distribution is constructed from the sum of two Crystal Ball functions, whose shape is determined from one-dimensional fits to the simulated signal decays. However, when fitting data a freely varying mass offset and resolution scaling factor are included to allow the \piz mass distribution to differ from that of the signal decays. No significant contribution from combinatorial decays with a real \etaz meson and an unrelated track is found when fitting $\decay{D_{(s)}^{+}}{h^{+}\etaz}$ decays, therefore no corresponding component is included. 

Decays of charm mesons to $h^{+}h^{0}X$ final states, where $X$ is at least one unreconstructed particle, appear as partially reconstructed background below the $D^{+}_{(s)}$ meson masses. Using external input on branching fractions and charm-meson production cross-section ratios~\mbox{\cite{Zyla:2020zbs,LHCb-PAPER-2015-041}} it is determined that only the decay \decay{\Dsp}{\pip\etaz\piz} has a significant contribution in the fit to $\decay{D^{+}_{(s)}}{\pip\etaz}$ candidates. 
To account for this component, a shape comprising an exponential function in the $m(h^{+}\etaz)$ dimension with a freely varying coefficient and a peaking $m(\ep\en\gamma)$ distribution constructed from two Crystal Ball functions is added. 

The fit to $\decay{D_{(s)}^{+}}{h^{+}\piz}$ ($\decay{D_{(s)}^{+}}{h^{+}\etaz}$) candidates includes 91 (54) freely varying parameters. The models are validated using pseudo-experiments and no significant biases in the values or statistical uncertainties of the raw asymmetries are observed.
The projections of the fits to $\decay{D_{(s)}^{+}}{h^{+}\piz}$ and $\decay{D_{(s)}^{+}}{h^{+}\etaz}$ candidates, summed over all relevant categories, are shown in Figs.~\ref{fig:results_pi0} and~\ref{fig:results_eta}, respectively. The pull distributions are examined for each category of the fit in both projections and in two dimensions, and no significant biases are seen. The goodness-of-fit is quantified by calculating the $\chi^{2}$ value for each projection and category separately, and combining to determine $\chi^{2}/N_{\text{dof}} = 0.90$ and $\chi^{2}/N_{\text{dof}} = 1.06$ for the fits to $\decay{D_{(s)}^{+}}{h^{+}\piz}$ and $\decay{D_{(s)}^{+}}{h^{+}\etaz}$ candidates, where $N_{\text{dof}}$ is the total number of degrees of freedom.
The corresponding signal yields and raw asymmetries are listed in Table~\ref{tab:raw_asym}.
\begin{table}[btp]
    \centering
    \caption{Signal yields in each running period and corresponding raw asymmetries for $\decay{D_{(s)}^{+}}{h^{+}\piz}$ and $\decay{D_{(s)}^{+}}{h^{+}\etaz}$ candidates. The uncertainties are statistical. }
    \begin{tabular}{c r@{\:$\pm$\:}l r@{\:$\pm$\:}l r@{\:$\pm$\:}l c }
    \toprule
        Mode & \multicolumn{6}{c}{Yield} &  $A_{\text{Raw}}$ (\%)  \\
        
             &  \multicolumn{2}{c}{2011} &  \multicolumn{2}{c}{2012} &  \multicolumn{2}{c}{Run 2} &  \\
        \midrule 
        \decay{\Dp}{\pip\piz} &740&	60	& 2\,240& 120 & 25\,750 &	430	& $-1.64 \pm \phantom{0}0.93$ \\
        \decay{\Dsp}{\pip\piz}& 20&	30	&  $-50$&  50 &	  450 &	120	& - \\
        \decay{\Dp}{\Kp\piz}  & 10&	13	&   90  &  30 &	2\,440 &	110	& $-2.53 \pm \phantom{0}4.75$ \\
        \decay{\Dsp}{\Kp\piz} & 54& 13	&   150	&  30 & 2\,580 &	90	& $-0.25 \pm \phantom{0}3.87$ \\
        \midrule 
        \decay{\Dp}{\pip\etaz}  &\multicolumn{2}{c}{-}&\multicolumn{2}{c}{-}&32\,760	&	380	& $-0.55 \pm \phantom{0}0.76$ \\
        \decay{\Dsp}{\pip\etaz} &\multicolumn{2}{c}{-}&\multicolumn{2}{c}{-}&37\,950	&	340	& $\phantom{+}0.75 \pm \phantom{0}0.65$   \\
        \decay{\Dp}{\Kp\etaz}   &\multicolumn{2}{c}{-}&\multicolumn{2}{c}{-}&880	&	70	& $-5.39 \pm 10.40$ \\
        \decay{\Dsp}{\Kp\etaz}  &\multicolumn{2}{c}{-}&\multicolumn{2}{c}{-}&2\,520	&	70	& $\phantom{+}1.28 \pm \phantom{0}3.67$   \\
        \bottomrule
    \end{tabular}
    \label{tab:raw_asym}
\end{table}
The \Dp and \Dsp signal distributions overlap, leading to small correlations between the measured raw asymmetries. The correlation coefficients are listed in Table~\ref{tab:correlations} and the largest correlation is 10\%.  
\begin{table}[btp]
    \centering
    \caption{Correlation coefficients between the raw asymmetries determined for $\decay{D_{(s)}^{+}}{h^{+}\piz}$ and $\decay{D_{(s)}^{+}}{h^{+}\etaz}$ decays. }
    \begin{tabular}{c c c c}
    \toprule
            &  \decay{\Dp}{\pip\piz}        &   \decay{\Dp}{\Kp\piz}        &  \decay{\Dsp}{\Kp\piz}        \\ 
    \midrule
    \decay{\Dp}{\pip\piz} & $\phantom{+}1.00$   &         & \\ 
    \decay{\Dp}{\Kp\piz}  & $-0.01$ & $1.00$  & \\
    \decay{\Dsp}{\Kp\piz} & $-0.09$ & $0.10$  & $\phantom{+}1.00$ \\
    \bottomrule
     & & & \\
    \end{tabular}\\
    
    \begin{tabular}{c c c c c }
    \toprule
            &  \decay{\Dp}{\pip\etaz} &   \decay{\Dp}{\Kp\etaz} & \decay{\Dsp}{\pip\etaz} &  \decay{\Dsp}{\Kp\etaz} \\ 
    \midrule
    \decay{\Dp}{\pip\etaz} & $\phantom{+}1.00$ &  &  & \\
    \decay{\Dp}{\Kp\etaz}  & $-0.00$ & $\phantom{+}1.00$ &  &  \\   
    \decay{\Dsp}{\pip\etaz}& $\phantom{+}0.01$  & $\phantom{+}0.00$ & $\phantom{+}1.00$ & \\ 
    \decay{\Dsp}{\Kp\etaz} & $-0.06$ & $\phantom{+}0.10$ & $-0.00$ & $\phantom{+}1.00$ \\
    \bottomrule
    \end{tabular}
    \label{tab:correlations}
\end{table}

\section{Control modes}
\label{sec:Control_Modes}
The impact of production and detection asymmetries of the signal modes is accounted for using large samples of $\decay{D_{(s)}^{+}}{\KS h^{+}}$ decays. 
The samples are selected using similar requirements to the signal modes, where possible. 
Candidates are built at the high-level software trigger stage by first combining two well-reconstructed hadronic tracks that are inconsistent with originating from any PV to create the \KS decay vertex. Similar to the electrons, these tracks must also have track segments within the \velo. The \KS candidate is combined with a hadronic track with either the pion or kaon mass hypothesis to form the $D^{+}_{(s)}$ decay vertex. The same momentum, pseudorapidity  and fiducial requirements are placed on the tracks as used for the signal.
The candidates are required to have $482 < m(\pip\pim) < 512 \mevcc$ and $1800 < m(\KS h^{+}) < 2050 \mevcc$, a proper decay time of $t > 0.25 \ps$, and the same direction angle and \pt requirements as the signal. 
Tighter PID requirements are placed on the control mode candidates than the signal to remove larger contamination from misidentification background.

\begin{figure}[btp]
  \begin{center}
    \includegraphics[width=0.49\linewidth]{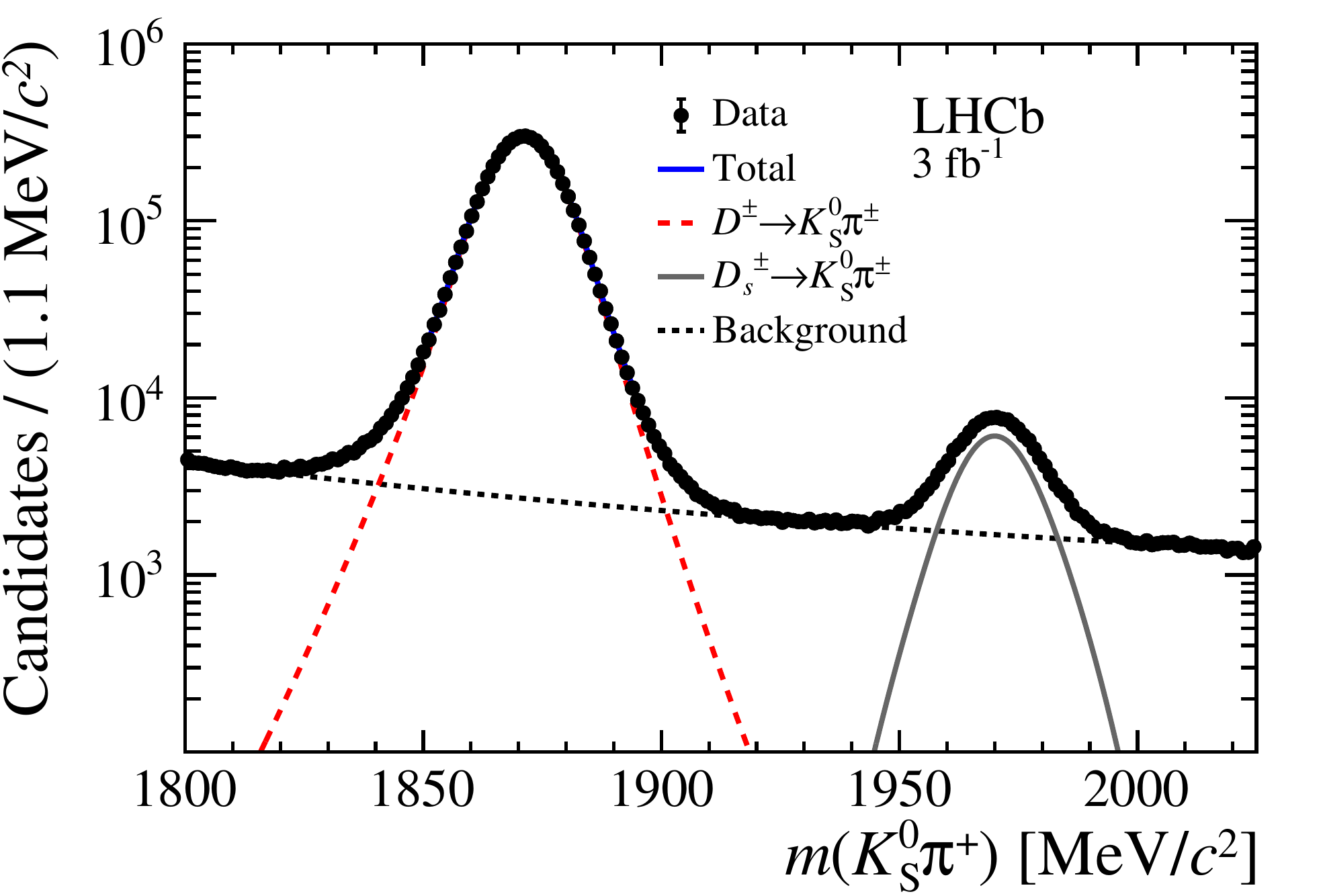}
    \includegraphics[width=0.49\linewidth]{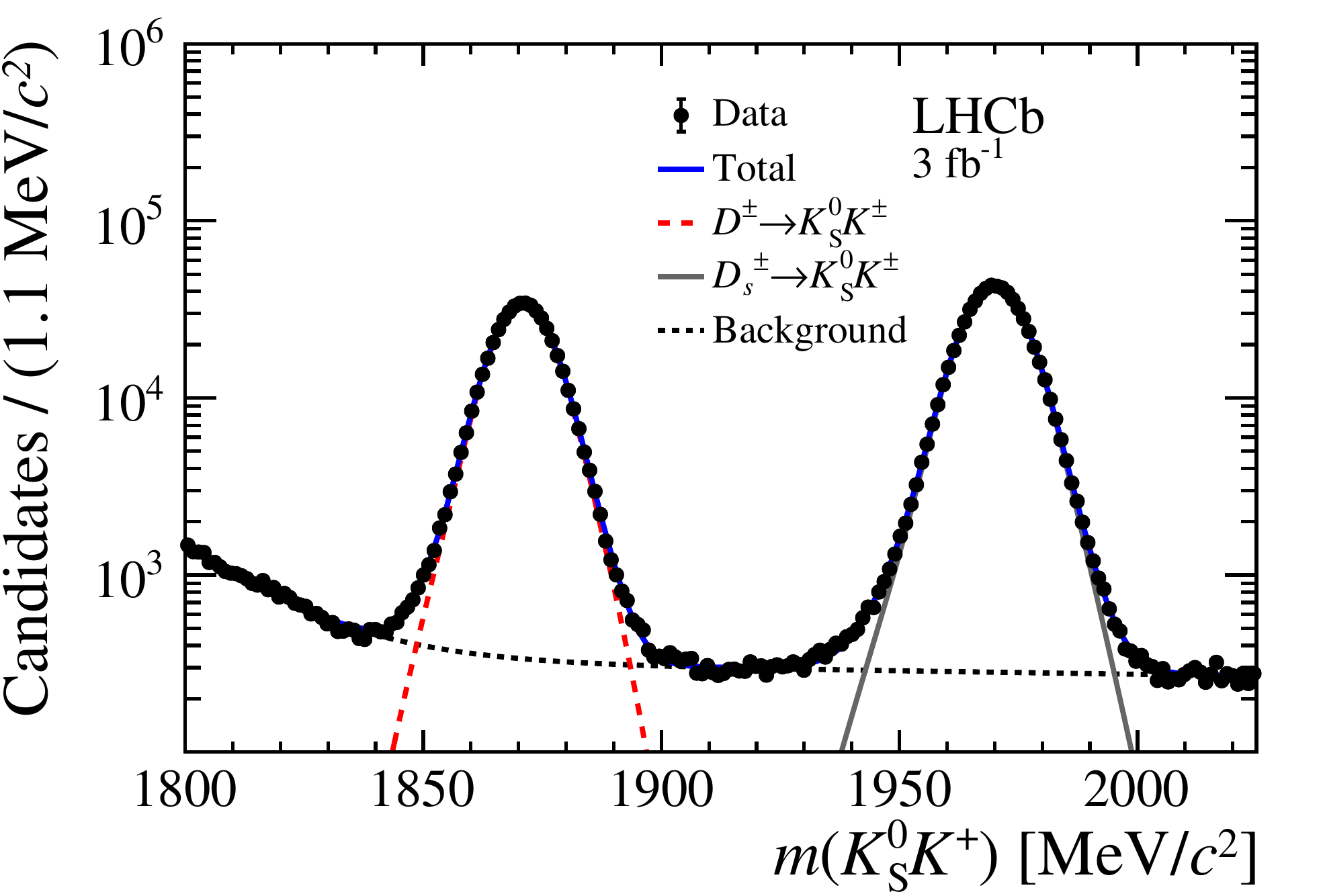}
    \includegraphics[width=0.49\linewidth]{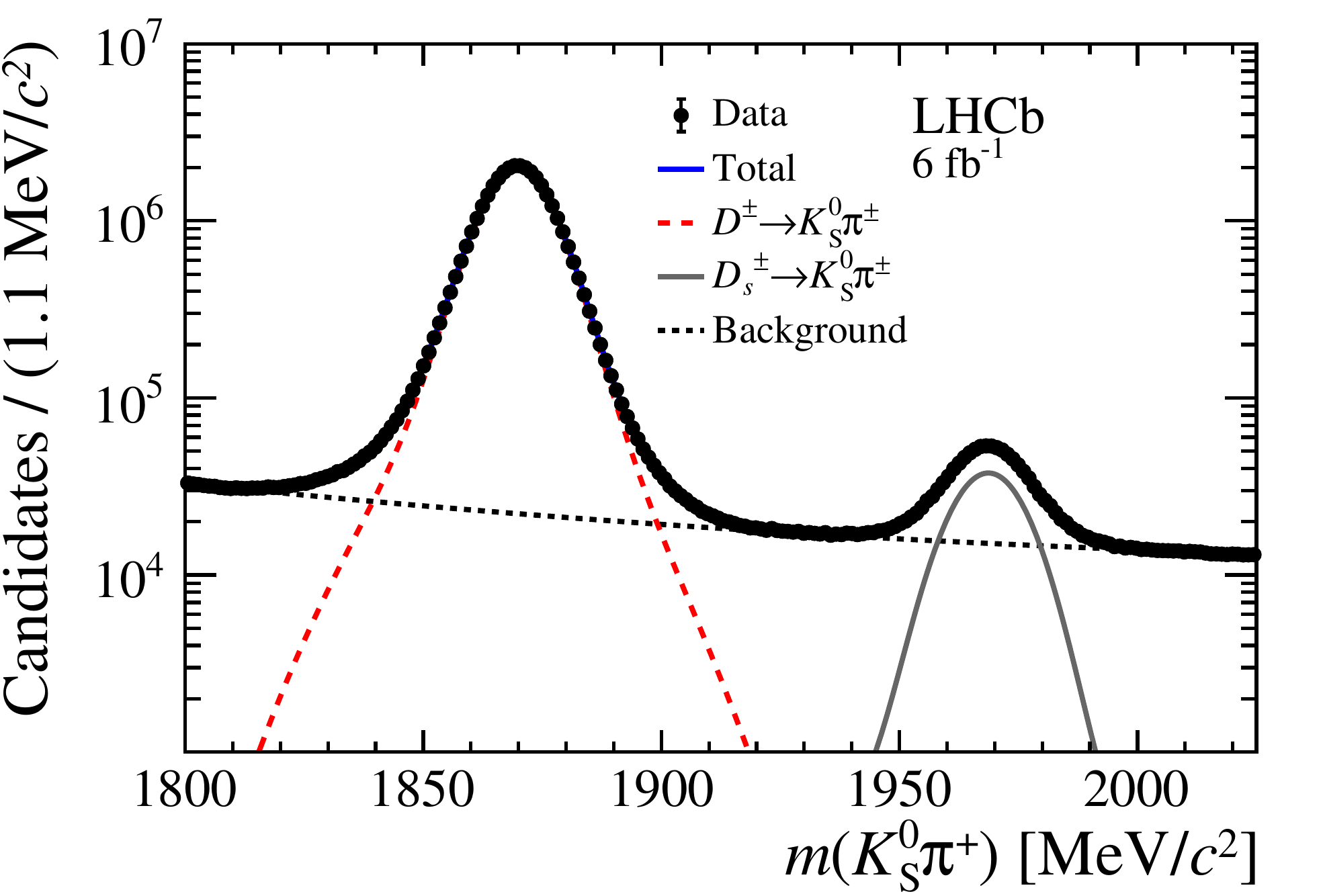}
    \includegraphics[width=0.49\linewidth]{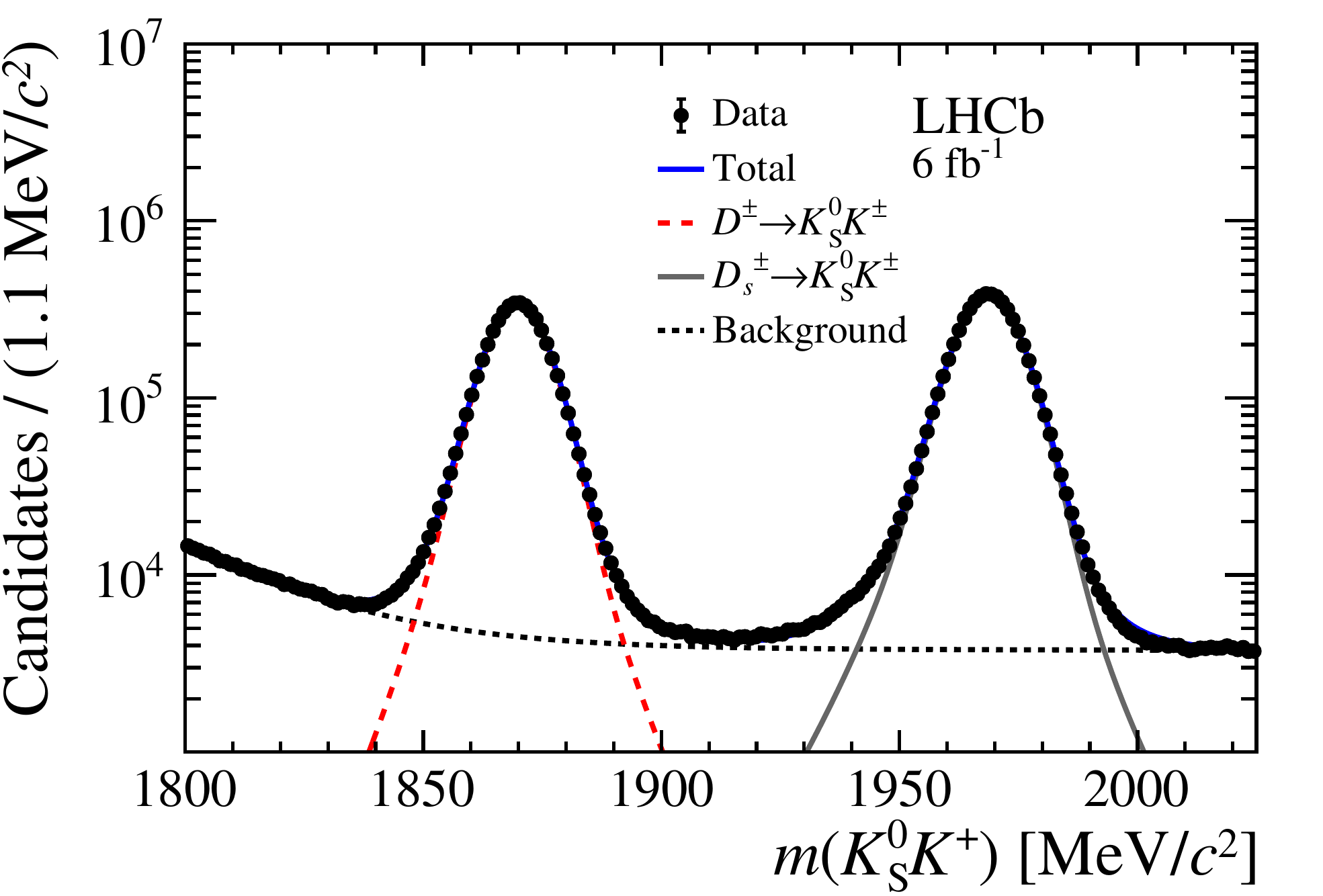}
  \end{center}
  \caption{
    Distributions of the (left) $m(\KS \pip)$ and (right) $m(\KS \Kp)$ mass  of control mode candidates in (top) Run 1 and (bottom) Run 2. The total PDF and individual fit components are overlaid, including $\decay{\Dp}{\KS h^{+}}$ decays in dashed red, $\decay{\Dsp}{\KS h^{+}}$ decays in solid grey and background decays in dashed black.
    }
  \label{fig:control_fits}
\end{figure}

The kinematic distributions of the signal and control candidates are determined using the \sPlot technique~\cite{Pivk:2004ty} with $m(\KS h^{+})$ as the discriminating variable for the latter. Binned maximum-likelihood fits are performed on the control-mode candidates using signal models comprising a Gaussian function and Johnson $S_{U}$ function~\cite{Johnson:1949zj} as described in Ref.~\cite{LHCb-PAPER-2019-002}. The results are shown in Fig.~\ref{fig:control_fits}. The weighting procedure is performed separately for Run~1 and Run~2 to allow for differences in the signal selection during these periods. To ensure the cancellation of the production and detection asymmetries, the relevant $D^{+}_{(s)}$ and $h^{+}$ kinematics (\ptot, azimuthal angle and pseudorapidity) are weighted to match those of the signal. Due to the large correlation between the $D^{+}_{(s)}$ and $h^{+}$ kinematics the weights for each variable are determined using a two-dimensional binning of the $D^{+}_{(s)}$ and $h^{+}$ distributions. In addition to the kinematics, weights are determined for the trigger category and IP distributions for the $D^{+}_{(s)}$ candidates. At the hardware trigger stage the candidates can be split into exclusive categories according to the origin of the positive trigger decision: the first category contains any candidate with a calorimeter deposit associated to the $h^{0}$ or \KS decay; the second category contains any remaining candidate with a deposit not associated to any of the signal particles; and the third category contains candidates still remaining with a high \pt deposit associated to the charged pion or kaon.
The control-mode candidates are weighted to reproduce the populations of signal candidates in each of these three categories. 

The IP of the $D_{(s)}^{+}$ candidate is indicative of whether the meson was produced in the primary interaction, or as a product of a \bquark-hadron decay, and therefore with a significant IP with respect to the PV, referred to as a secondary decay. In the latter case the production asymmetry of the parent \bquark-hadron could differ from that of the \Dp or \Dsp meson. The signal and control mode selections require that the $D^{+}_{(s)}$ candidates are consistent with originating at a PV, suppressing the fraction of candidates from secondary decays to less than 10\%. 
If the fraction of $D_{(s)}^{+}$ candidates from the primary interaction and secondary decays varies between the signal and control mode then the production asymmetries may not exactly cancel, therefore the control sample is weighted to match the IP distribution of the signal.    

Binned maximum-likelihood fits are performed to the charge-split samples to determine the raw asymmetries separately for Run~1 and Run~2.  The signals are described using the sum of a Gaussian function and Johnson $S_{U}$ function, using the same model as described in Ref.~\cite{LHCb-PAPER-2019-002}. The fits are performed after the samples have been weighted to match the kinematics of the signal modes, and the statistical uncertainty is calculated using the weights to account for the loss of precision resulting from the weighting procedure.

\section{Systematic uncertainties}
\label{sec:Systematic_Uncertainties}

The systematic uncertainty on the \CP asymmetries receives contributions from a number of sources, including the signal and background parameterisations, the control modes and selection requirements. 
\begin{table}[btp]
    \centering
    \caption{Absolute systematic uncertainties (\%) on the \CP asymmetries for $\decay{\D^{+}_{(s)}}{h^+\piz}$ decays.}
    \begin{tabular*}{\textwidth}{@{~\extracolsep{\fill}} l c   c  c}
         \toprule 
         Source & $\decay{\Dp}{\pip\piz}$ & $\decay{\Dp}{\Kp\piz}$ & $\decay{\Dsp}{\Kp\piz}$ \\
         \midrule
        Fit model                           
        &  $\phantom{<\,}0.59$  & $\phantom{<\,}1.55$ & $\phantom{<\,}1.01$ \\ 
        PID asymmetry                       
        &  $\phantom{<\,}0.06$  & $\phantom{<\,}0.27$ & $\phantom{<\,}0.15$ \\ 
        Secondary decays                    
        & $<0.01$& $\phantom{<\,}0.01$ & $\phantom{<\,}0.02$ \\ 
        Combined $A_{\text{Raw}}$ Run 1 and Run 2
        & $\phantom{<\,}0.23$ & $\phantom{<\,}0.65$ & $\phantom{<\,}0.30$ \\ 
        Control modes                       
        & $\phantom{<\,}0.03$	& $\phantom{<\,}1.18$	& $\phantom{<\,}0.59$	\\
        $A_{\text{Mix}}(K^0)$               
        & $<0.01$ & $<0.01$ & $<0.01$ \\ 
        $\mathcal{A}_{\CP}(\decay{D_{(s)}^{+}}{\KS h^{+}})$              
        & $\phantom{<\,}0.12$ & $\phantom{<\,}0.08$ & $\phantom{<\,}0.26$ \\ 
        \midrule
        Total                               
        & $\phantom{<\,}0.65$ & $\phantom{<\,}2.07$ & $\phantom{<\,}1.24$ \\
        \bottomrule
        \end{tabular*}
    \label{tab:total_systematics_pi0}
\end{table}
\begin{table}[btp]
    \centering
    \caption{Absolute systematic uncertainties (\%) on the \CP asymmetries for $\decay{\D^{+}_{(s)}}{h^+\etaz}$ decays.}
    \begin{tabular*}{\textwidth}{@{~\extracolsep{\fill}} l c   c  c c }
         \toprule
         Source & $\decay{\Dp}{\pip\eta}$ &$\decay{\Dsp}{\pip\eta}$ & $\decay{\Dp}{\Kp\eta}$ & $\decay{\Dsp}{\Kp\eta}$\\
         \midrule
        Fit model                           
        & $\phantom{<\,}0.35$ & $\phantom{<\,}0.15$ & $\phantom{<\,}4.04$ & $\phantom{<\,}1.08$ \\ 
        PID asymmetry                       
        & $\phantom{<\,}0.06$ & $\phantom{<\,}0.01$ & $\phantom{<\,}0.87$ & $\phantom{<\,}0.16$ \\ 
        Secondary decays                    
        & $< 0.01$ & $\phantom{<\,}0.02$ & $\phantom{<\,}0.01$ & $\phantom{<\,}0.04$ \\ 
        Control modes                       
        & $\phantom{<\,}0.05$ & $\phantom{<\,}0.39$  & $\phantom{<\,}0.14$	& $\phantom{<\,}0.12$ \\ 
        $A_{\text{Mix}}(K^0)$               
        & $< 0.01$ & $<0.01$ & $<0.01$ & $<0.01$ \\ 
        $\mathcal{A}_{\CP}(\decay{D_{(s)}^{+}}{\KS h^{+}})$              
        & $\phantom{<\,}0.12$ & $\phantom{<\,}0.20$ & $\phantom{<\,}0.08$ & $\phantom{<\,}0.26$ \\ 
        \midrule
        Total                               
        & $\phantom{<\,}0.38$ & $\phantom{<\,}0.46$ & $\phantom{<\,}4.13$ & $\phantom{<\,}1.13$ \\
        \bottomrule
        \end{tabular*}
    \label{tab:total_systematics_eta}
\end{table}
The assumptions used when creating the signal and background parameterisations are varied and the corresponding systematic uncertainty is quantified using the resulting difference in the raw asymmetries in the fits to data. This includes using: alternative signal parameterisation comprising Johnson $S_{U}$ functions instead of Crystal Ball functions; different pure-combinatorial $m(h^{+}h^{0})$ parameterisations of a constant plus exponential function;  alternative pure-combinatorial $m(\ep\en\gamma)$ parameterisations of a third-order Chebychev polynomial function; alternative real-\piz combinatorial parameterisation of a double Johnson $S_{U}$; and different misidentification-background parameterisations using Johnson $S_{U}$ functions instead of Crystal Ball functions. The efficiencies used to constrain the level of misidentification background are varied within the corresponding uncertainties in 100 fits and the spread in the raw asymmetries is used to estimate the systematic uncertainty. The impact on the raw asymmetries is quantified when various neglected background components are included in the model, including semileptonic $\decay{D^{+}_{(s)}}{h^{0}\ep\neue}$ and $\decay{D^{+}_{(s)}}{h^{0}\mup\neum}$ decays, partially reconstructed $\decay{\Dz}{\Km\pip\piz}$ decays and a combinatorial component with a real-\etaz distribution. Additionally, the assumption that the pure-combinatorial $m(h^{+}h^{0})$ exponential slope is independent of $m(\ep\en\gamma)$ is relaxed by allowing a linear dependence. The signal tail parameters that are fixed to values obtained from simulation are allowed to vary with an overall scaling factor and the impact on the raw asymmetries is quantified. The assumption that the mean $D^{+}_{(s)}$ mass positions are the same for $D_{(s)}^{+}$ and $D_{(s)}^{-}$ candidates is tested by allowing different values. 
The systematic uncertainty from the fit model is dominated by the fixed tail parameters for $\decay{\Dp}{\pip\piz}$, the fixed misidentification efficiency ratio for $\decay{\Dp}{\Kp\piz}$ and $\decay{\Dsp}{\Kp\piz}$ decays, the signal parameterisation for $\decay{\Dp}{\pip\etaz}$ decays and the lack of real-\etaz combinatorial contribution for $\decay{\Dsp}{\pip\etaz}$, $\decay{\Dp}{\Kp\etaz}$ and $\decay{\Dsp}{\Kp\etaz}$ decays. 

The selection of the signal and control modes uses different requirements for the PID variables. Tighter conditions are needed for the control modes to reduce misidentification background such as \decay{\Lc}{\proton\KS} decays. The size of a possible charge asymmetry induced by these different requirements is quantified by first computing the asymmetry of the PID efficiencies, $\epsilon_{\text{PID}}$, when determined separately for positively and negatively charged hadrons, $A_{\text{PID}} = [\epsilon_{\text{PID}}(h^{+})-\epsilon_{\text{PID}}(h^{-})]/[\epsilon_{\text{PID}}(h^{+})+\epsilon_{\text{PID}}(h^{-})]$. Then, the difference in PID asymmetry when calculated using signal and control mode PID requirements, $\Delta A_{\text{PID}} = A_{\text{PID}}^{\text{signal}} - A_{\text{PID}}^{\text{control}}$, is used to quantify the corresponding systematic uncertainty.
Additionally, the difference in the raw asymmetries when not performing the IP weighting is used to quantify the systematic uncertainty arising from the secondary decays. 

The asymmetries for $\decay{\D^{+}_{(s)}}{h^+\piz}$ decays are determined from simultaneous fits to data sets taken during Run 1 and Run 2, with a single \CP asymmetry shared between the categories for each mode. In contrast, the control-mode fits are performed separately for Run 1 and Run 2 and then a weighted average is performed to combine the measurements, where the weighting is determined from the yields of signal mode decays. The systematic uncertainty arising from this method is quantified by performing the signal fits separately for Run 1 and Run 2, taking the appropriate difference with the control-mode asymmetries and then combining the Run 1 and Run 2 results to get an alternative estimate. 

The control-mode weighting is performed in nearly equally populated bins. The binning scheme is varied to determine the associated systematic uncertainty. After performing the weighting procedure, the remaining discrepancies in the kinematic distributions are quantified by summing the difference in the normalised distributions of signal and control modes, multiplied by the local asymmetry minus the average asymmetry. The fit model used to measure the control-mode raw asymmetries is varied from the sum of a Johnson $S_{U}$ function and a Gaussian function to the sum of a Crystal Ball function and a Gaussian function. The contribution to the control mode raw asymmetry from the neutral-kaon mixing and decay asymmetry is calculated and the corresponding uncertainty of this calculation is dominated by the knowledge of the detector material. The uncertainties of the external values of the control mode $\mathcal{A}_{\CP}$ are included as systematic uncertainties.

The systematic uncertainties are listed for the $\decay{\D^{+}_{(s)}}{h^+\piz}$ modes in Table~\ref{tab:total_systematics_pi0} and for the $\decay{\D^{+}_{(s)}}{h^+\etaz}$ modes in Table~\ref{tab:total_systematics_eta}. These are dominated by the fit-model uncertainty in most cases, except for the mode $\decay{\Dsp}{\pip\etaz}$ which is dominated by the uncertainty arising from the control mode $\decay{\Dsp}{\KS\pip}$, the smallest of the control samples.

As a crosscheck, the fits are performed in various subsamples: split by year of data taking; magnet polarity; trigger category; bremsstrahlung category; $D^{+}_{(s)}$ kinematics and $h^{+}$ kinematics. No significant biases are found with respect to the nominal results. 

\section{Results and conclusions}
\label{sec:Results_and_conc}

The \CP asymmetries are calculated using Eq.~\ref{eq:result},
where for each mode the corresponding control channel $A_{\text{Raw}}^{\text{w}}$, independently measured $\mathcal{A}_{\CP}(\decay{D_{(s)}^{+}}{\KS h^{+}})$ and calculated $A_{\text{Mix}}(K^0)$ are taken. The final results are listed in Tables~\ref{tab:results_pi0} and \ref{tab:results_eta}. 
The results are shown with the corresponding statistical uncertainty from the fits and the total systematic uncertainty as listed in Tables~\ref{tab:total_systematics_pi0} and \ref{tab:total_systematics_eta}. The systematic uncertainties attributed to $A_{\text{Raw}}^{\text{w}}(\decay{D_{(s)}^{+}}{\KS h^{+}})$, $\mathcal{A}_{\CP}(\decay{D_{(s)}^{+}}{\KS h^{+}}) $ and $A_{\text{Mix}}(K^0)$ are listed separately. 

\begin{table}[btp]
    \centering
    \caption{Final $\mathcal{A}_\CP$ (\%) results for the $\decay{D_{(s)}^{+}}{h^{+}\piz}$ modes. The uncertainties of $\mathcal{A}_{\CP}(\decay{D_{(s)}^{+}}{h^{+}\piz})$ are statistical and systematic respectively. The uncertainties of $A_{\text{Raw}}(\decay{D_{(s)}^{+}}{h^{+}\piz})$ are purely statistical. The uncertainties of $A_{\text{Mix}}(K^0)$ are systematic. Externally measured values of $\mathcal{A}_{\CP}(\decay{D_{(s)}^{+}}{\KS h^{+}})$ are taken from Refs.~\cite{Ko:2012pe,LHCb-PAPER-2019-002,Ablikim:2019whl,Lees:2012jv,Onyisi:2013bjt,Ko:2010ng}. For comparison the unweighted control asymmetries are  $\mathcal{A}_{\text{Raw}}(\decay{\Dp}{\KS\pip}) = -0.45 \pm 0.02$, 
    $\mathcal{A}_{\text{Raw}}(\decay{\Dp}{\KS\Kp}) = 0.47 \pm 0.05$ and
    $\mathcal{A}_{\text{Raw}}(\decay{\Dsp}{\KS\Kp}) = 0.51 \pm 0.04$.}
    \begin{tabular*}{\textwidth}{@{~\extracolsep{\fill}}  l  c  c  c  }
        \toprule
        & \decay{\Dp}{\pip\piz} & \decay{\Dp}{\Kp\piz}  & \decay{\Dsp}{\Kp\piz} \\
        \midrule
        $A_{\text{Raw}}(\decay{D_{(s)}^{+}}{h^{+}\piz})$			        & $-1.64\phantom{0} \pm 0.93\phantom{0}$	& $-2.53\phantom{0} \pm 4.75\phantom{0}$	& $-0.25\phantom{0} \pm 3.87\phantom{0}$ \\
        $A_{\text{Raw}}^{\text{w}}(\decay{D_{(s)}^{+}}{\KS h^{+}})$	& $-0.45\phantom{0} \pm 0.02\phantom{0}$	& $\phantom{+}0.58\phantom{0} \pm 0.08\phantom{0}$	& $\phantom{+}0.60\phantom{0} \pm 0.07\phantom{0}$ \\
        $\mathcal{A}_{\CP}(\decay{D_{(s)}^{+}}{\KS h^{+}})$			        & $-0.02\phantom{0} \pm 0.12\phantom{0}$	& $-0.01\phantom{0} \pm 0.08\phantom{0}$	& $\phantom{+}0.09\phantom{0} \pm 0.26\phantom{0}$ \\
        $A_{\text{Mix}}(K^0)$							                    & $-0.070 \pm 0.004$	& $-0.072 \pm 0.004$	& $-0.072 \pm 0.004$ \\
        \midrule
        $\mathcal{A}_{\CP}(\decay{D_{(s)}^{+}}{h^{+}\piz})$				    & $-1.3 \pm 0.9 \pm 0.6$	& $-3.2 \pm 4.7 \pm 2.1$	& $-0.8 \pm 3.9 \pm 1.2$ \\
        \bottomrule
    \end{tabular*}
    \label{tab:results_pi0}
\end{table}
\begin{table}[btp]
    \centering
    \caption{Final $\mathcal{A}_\CP$ (\%) results for the $\decay{D_{(s)}^{+}}{h^{+}\etaz}$ modes. The uncertainties of $\mathcal{A}_{\CP}(\decay{D_{(s)}^{+}}{h^{+}\etaz})$ are statistical and systematic respectively. The uncertainties of $A_{\text{Raw}}(\decay{D_{(s)}^{+}}{h^{+}\etaz})$ are purely statistical. The uncertainties of $A_{\text{Mix}}(K^0)$ are systematic. Externally measured values of $\mathcal{A}_{\CP}(\decay{D_{(s)}^{+}}{\KS h^{+}})$ are taken from Refs.~\cite{Ko:2012pe,LHCb-PAPER-2019-002,Ablikim:2019whl,Lees:2012jv,Onyisi:2013bjt,Ko:2010ng}. For comparison the unweighted control asymmetries are  $\mathcal{A}_{\text{Raw}}(\decay{\Dp}{\KS\pip}) = -0.45 \pm 0.02$, 
    $\mathcal{A}_{\text{Raw}}(\decay{\Dsp}{\KS\pip}) = -0.13 \pm 0.17$, 
    $\mathcal{A}_{\text{Raw}}(\decay{\Dp}{\KS\Kp}) = 0.47 \pm 0.05$ and
    $\mathcal{A}_{\text{Raw}}(\decay{\Dsp}{\KS\Kp}) = 0.51 \pm 0.04$.}
    \begin{tabular*}{1\textwidth}{@{~\extracolsep{\fill}}  l  c  c  }
        \toprule
        & \decay{\Dp}{\pip\etaz} & \decay{\Dsp}{\pip\etaz}\\
        \midrule
        $A_{\text{Raw}}(\decay{D_{(s)}^{+}}{h^{+}\etaz})$			& $-0.55\phantom{0} \pm 0.76\phantom{0}$	& $\phantom{+}0.75\phantom{0} \pm 0.65\phantom{0}$\\
        $A_{\text{Raw}}^{\text{w}}(\decay{D_{(s)}^{+}}{\KS h^{+}})$	& $-0.46\phantom{0} \pm 0.04\phantom{0}$	& $-0.02\phantom{0} \pm 0.37\phantom{0}$	\\
        $\mathcal{A}_{\CP}(\decay{D_{(s)}^{+}}{\KS h^{+}})$				& $-0.02\phantom{0} \pm 0.12\phantom{0}$	& $\phantom{+}0.13\phantom{0} \pm 0.20\phantom{0}$ \\
        $A_{\text{Mix}}(K^0)$							                & $-0.070 \pm 0.004$	& $-0.070 \pm 0.004$\\
        \midrule
        $\mathcal{A}_{\CP}(\decay{D_{(s)}^{+}}{h^{+}\etaz})$		    & $-0.2 \pm 0.8 \pm 0.4$	& $0.8 \pm 0.7 \pm 0.5$\\    
        \bottomrule
    \end{tabular*}\\

    \begin{tabular*}{1\textwidth}{@{~\extracolsep{\fill}}  l  c  c  }
        &&\\
        \toprule
        & \decay{\Dp}{\Kp\etaz}  & \decay{\Dsp}{\Kp\etaz}\\
        \midrule
        $A_{\text{Raw}}(\decay{D_{(s)}^{+}}{h^{+}\etaz})$			    & $-5.39\phantom{0} \pm 10.40\phantom{0}$	& $\phantom{+}1.28\phantom{0} \pm 3.67\phantom{0}$ \\
        $A_{\text{Raw}}^{\text{w}}(\decay{D_{(s)}^{+}}{\KS h^{+}})$	& $\phantom{+}0.33\phantom{0} \pm \phantom{0}0.10\phantom{0}$	    & $\phantom{+}0.36\phantom{0} \pm 0.10\phantom{0}$ \\
        $\mathcal{A}_{\CP}(\decay{D_{(s)}^{+}}{\KS h^{+}})$				& $-0.01\phantom{0} \pm \phantom{0}0.08\phantom{0}$	& $\phantom{+}0.09\phantom{0} \pm 0.26\phantom{0}$ \\
        $A_{\text{Mix}}(K^0)$							                & $-0.073 \pm \phantom{0}0.004$	& $-0.073 \pm 0.004$ \\
        \midrule
        $\mathcal{A}_{\CP}(\decay{D_{(s)}^{+}}{h^{+}\etaz})$			& $-6 \pm 10 \pm 4$	& $0.9 \pm 3.7 \pm 1.1$ \\    
        \bottomrule
    \end{tabular*}
    \label{tab:results_eta}
\end{table}

In summary, measurements of \CP asymmetries in $\decay{D_{(s)}^{+}}{h^{+}\piz}$ and $\decay{D_{(s)}^{+}}{h^{+}\etaz}$ decays are performed using \proton\proton collision data corresponding to 9\invfb and 6\invfb of integrated luminosity collected at the \lhcb experiment, respectively. The neutral mesons are reconstructed using the $\ep\en\gamma$ final state, allowing the $D^{+}_{(s)}$ decay vertex to be reconstructed. 
The production and detection asymmetries are cancelled using large samples of $\decay{D_{(s)}^{+}}{\KS h^{+}}$ decays, weighted to match the kinematics of the signal modes. 
The \CP asymmetries are determined to be 
\begin{alignat*}{7}
    \mathcal{A}_{\CP}(\decay{\Dp}{\pip\piz}) 	&= (-&&1.3 &&\pm 0.9 &&\pm 0.6 &)\%, \\
    \mathcal{A}_{\CP}(\decay{\Dp}{\Kp\piz}) 	&= (-&&3.2 &&\pm 4.7 &&\pm 2.1 &)\%, \\
    \mathcal{A}_{\CP}(\decay{\Dp}{\pip\etaz})   &= (-&&0.2 &&\pm 0.8 &&\pm 0.4 &)\%, \\
    \mathcal{A}_{\CP}(\decay{\Dp}{\Kp\etaz}) 	&= (-&&6 &&\pm 10 &&\pm 4 &)\%, \\
    \mathcal{A}_{\CP}(\decay{\Dsp}{\Kp\piz}) 	&= (-&&0.8 &&\pm 3.9 &&\pm 1.2 &)\%, \\
    \mathcal{A}_{\CP}(\decay{\Dsp}{\pip\etaz})  &= (&&0.8 &&\pm 0.7 &&\pm 0.5 &)\%, \\
    \mathcal{A}_{\CP}(\decay{\Dsp}{\Kp\etaz})   &= (&&0.9 &&\pm 3.7 &&\pm 1.1 &)\%,
\end{alignat*}
where the first uncertainty is statistical and the second systematic. 
All of the results are consistent with no \CP asymmetry and the first five 
constitute the most precise measurements to date. Very recently 
the Belle collaboration has also reported precise measurements of $\mathcal{A}_{\CP}(\decay{\Dsp}{\Kp\piz})$, $\mathcal{A}_{\CP}(\decay{\Dsp}{\pip\etaz})$ and $\mathcal{A}_{\CP}(\decay{\Dsp}{\Kp\etaz})$~\cite{1852254}.
The result for $\mathcal{A}_{\CP}(\decay{\Dp}{\pip\piz})$ is consistent with the SM expectation and the previous measurement by the Belle collaboration~\cite{Babu:2017bjn}. Using the relevant lifetimes, branching fractions and \CP asymmetries from Ref.~\cite{Zyla:2020zbs} and an updated average of $\mathcal{A}_{\CP}(\decay{\Dp}{\pip\piz}) = (0.43 \pm 0.79)\%$ calculated using the measurements by Belle~\cite{Babu:2017bjn}, CLEO~\cite{Mendez:2009aa} and the result presented here, the isospin sum rule defined in Eq.~\ref{eq:sumrule} is found to be consistent with zero, with a value of $R=(0.1 \pm 2.4)\times10^{-3}$.

%% file: acknowledgements.tex
\section*{Acknowledgements}
%
%
\noindent We express our gratitude to our colleagues in the CERN
accelerator departments for the excellent performance of the LHC. We
thank the technical and administrative staff at the LHCb
institutes.
We acknowledge support from CERN and from the national agencies:
CAPES, CNPq, FAPERJ and FINEP (Brazil); 
MOST and NSFC (China); 
CNRS/IN2P3 (France); 
BMBF, DFG and MPG (Germany); 
INFN (Italy); 
NWO (Netherlands); 
MNiSW and NCN (Poland); 
MEN/IFA (Romania); 
MSHE (Russia); 
MICINN (Spain); 
SNSF and SER (Switzerland); 
NASU (Ukraine); 
STFC (United Kingdom); 
DOE NP and NSF (USA).
We acknowledge the computing resources that are provided by CERN, IN2P3
(France), KIT and DESY (Germany), INFN (Italy), SURF (Netherlands),
PIC (Spain), GridPP (United Kingdom), RRCKI and Yandex
LLC (Russia), CSCS (Switzerland), IFIN-HH (Romania), CBPF (Brazil),
PL-GRID (Poland) and NERSC (USA).
We are indebted to the communities behind the multiple open-source
software packages on which we depend.
Individual groups or members have received support from
ARC and ARDC (Australia);
AvH Foundation (Germany);
EPLANET, Marie Sk\l{}odowska-Curie Actions and ERC (European Union);
A*MIDEX, ANR, Labex P2IO and OCEVU, and R\'{e}gion Auvergne-Rh\^{o}ne-Alpes (France);
Key Research Program of Frontier Sciences of CAS, CAS PIFI, CAS CCEPP, 
Fundamental Research Funds for the Central Universities, 
and Sci. \& Tech. Program of Guangzhou (China);
RFBR, RSF and Yandex LLC (Russia);
GVA, XuntaGal and GENCAT (Spain);
the Leverhulme Trust, the Royal Society
 and UKRI (United Kingdom).

%% file: Authorship_LHCb-PAPER-2021-001.tex
\centerline
{\large\bf LHCb collaboration}
\begin
{flushleft}
\small
R.~Aaij$^{32}$,
C.~Abell{\'a}n~Beteta$^{50}$,
T.~Ackernley$^{60}$,
B.~Adeva$^{46}$,
M.~Adinolfi$^{54}$,
H.~Afsharnia$^{9}$,
C.A.~Aidala$^{85}$,
S.~Aiola$^{25}$,
Z.~Ajaltouni$^{9}$,
S.~Akar$^{65}$
% ...

{\footnotesize \it

% ...
$^{9}$Universit{\'e} Clermont Auvergne, CNRS/IN2P3, LPC, Clermont-Ferrand, France\\
% ...
$^{25}$INFN Sezione di Milano, Milano, Italy\\
% ...
$^{32}$Nikhef National Institute for Subatomic Physics, Amsterdam, Netherlands\\
% ...
$^{46}$Instituto Galego de F{\'\i}sica de Altas Enerx{\'\i}as (IGFAE), Universidade de Santiago de Compostela, Santiago de Compostela, Spain\\
% ...
$^{50}$Physik-Institut, Universit{\"a}t Z{\"u}rich, Z{\"u}rich, Switzerland\\
% ...
$^{54}$H.H. Wills Physics Laboratory, University of Bristol, Bristol, United Kingdom\\
% ...
$^{60}$Oliver Lodge Laboratory, University of Liverpool, Liverpool, United Kingdom\\
% ...
$^{65}$University of Cincinnati, Cincinnati, OH, United States\\
% ...
$^{68}$Syracuse University, Syracuse, NY, United States\\
% ...
$^{85}$University of Michigan, Ann Arbor, United States, associated to $^{68}$\\
\bigskip
% ...
\medskip
}
\end{flushleft}